\documentclass[aps,preprint]{revtex4}
\usepackage{amsfonts}
\usepackage{amssymb}
\usepackage{amsmath}
\usepackage{graphicx}
\usepackage[font={footnotesize,it}]{caption}
\usepackage{bigints}
\usepackage{latexsym}
\usepackage{enumitem}
\usepackage{hyperref}
\usepackage{xcolor}

\begin{document}
\title{On thermal fluctuations and quantum regularities of $\mathcal{F}(R,\mathcal{G})$ gravity black holes with constant topological Euler density in nonlinear electrodynamics}

\author{Mert Mangut}
\email{mert.mangut@emu.edu.tr}
\affiliation{Department of Physics, Eastern Mediterranean
University, 99628, Famagusta, North Cyprus via Mersin 10, Turkey}
\author{\"{O}zay G\"{u}rtu\u{g}}
\email{ozaygurtug@maltepe.edu.tr}
\affiliation{T. C. Maltepe University, Faculty of Engineering and Natural Sciences,
34857, Istanbul -Turkey}
\author{\.{I}zzet Sakall{\i}}
\email{izzet.sakalli@emu.edu.tr}
\affiliation{AS245, Department of Physics, Eastern Mediterranean
University, 99628, Famagusta, North Cyprus via Mersin 10, Turkey}

\begin{abstract}
We discuss the corrected thermodynamics and naked singularity structure of the topological static spherically symmetric solution in  $\mathcal{F}(R,\mathcal{G})$ - gravity coupled with Born-Infeld - like nonlinear electrodynamics. Solutions admitting black holes with constant topological Euler density is analyzed in view of various thermodynamical variables. The inclusion of logarithmic correction to the entropy is extended to the other thermodynamical variables and the contribution of corrected variables are displayed on various plots. The stability of black hole under the effect of thermal variables is also studied. As a second scope of this study, solutions admitting timelike naked singularity are probed with bosonic and fermionic quantum wave packets to see if the singularity is quantum mechanically regular or not. In this context, the evolution of these probes remains well-defined if the corresponding spatial Hamiltonian operator is essentially self-adjoint. Our calculations reveal that when the singularity is probed with specific wave modes involving spin - $0$ and spin - $1/2$  quantum wave packets, the corresponding wave operators turn out to be essentially self-adjoint, which in turn implies unique well-defined time evolution.
\end{abstract}

\maketitle

\section{Introduction}
Despite the outstanding accomplishments of Einstein's classical theory of general relativity, there are fundamental problems that make this theory inadequate to fully comprehend the physics of the cosmos. Dark matter and dark energy, which have been related to the expansion of the universe and spacetime singularities, maybe the most fundamental examples of these problems among others. Understanding and resolving these concepts require new laws of physics. The equations of classical general relativity are rendered useless by the high spacetime curvature present close to the singularity, necessitating the development of a new theory. Although quantum gravity seems to be a promising candidate when one wishes to investigate phenomena at the scale of singularities, such a consistent theory has still not been developed.
\\
Einstein's classical theory of general relativity can be extended in such a way that higher-order curvature terms are taken into account in addition to the Einstein-Hilbert term, in conjunction with the higher-dimensional extension of the classical theory of general relativity, to address the acceleration of the universe. The $\mathcal{F}(R)$ modified gravity theory is a well-known extension of the classical general relativity in which the standard Einstein's gravity is modified with an arbitrary function of the Ricci scalar ($R$) \citep{1,2}. The modified Gauss-Bonnet gravity model \big($\mathcal{F}(\mathcal{G})$ - gravity\big) inspired by string theory is another alternative theory that is found to be useful for explaining late-time acceleration of the universe \citep{3,4}. In addition to the aforementioned models, the $\mathcal{F}(R,\mathcal{G})$ - gravity, which admits topological static spherically symmetric solutions, can also be considered as an interesting model, as it incorporates both $R$ and the Gauss-Bonnet invariant \citep{5}. The matter field extension to the solution given in \citep{5} is considered in \citep{6}. In this study, a class of four-dimensional static spherically symmetric geometries with constant topological Euler density is studied when coupled with models of Born-Infeld-like nonlinear electrodynamics (NLED). Although the properties of black hole solutions obtained both in \citep{5} and \citep{6} have been analyzed in a certain extent, they deserve more detailed analysis by taking the corrected thermodynamics variables into account. The other important point that should be investigated is the solution admitting a time-like naked singularity, which poses a threat to the cosmic censorship hypothesis. To our knowledge, the structure of this singularity has been investigated neither in \citep{5, 6} nor in elsewhere. \\
The purpose of this study is two-fold. In the first part, we aim to investigate the thermal properties of the black hole solutions in view of corrected thermodynamic variables. This will be accomplished by taking the quantum fluctuation effects into account. \\
{\color{black} One of the main objectives of quantum gravity is to understand the nature of the microscopic degrees of freedom of black holes. The Boltzmann entropy formula:
\begin{equation}
S=k_B \ln \Omega_m,  \label{is1}  
\end{equation}
is a vital tool for this, where $k_B$ is the Boltzmann constant and $\Omega_m$ is the number of microstates in the system. The proper matching of microscopic and macroscopic computations of black hole entropy constitutes a crucial test of any quantum theory of gravity \cite{izs1,izs2,izs3,izs4}. Due to ongoing random transitions between its microstates, a system in contact with a thermal reservoir often experiences energy fluctuations around its equilibrium value. It is already well-known that such energy fluctuations for non-extremal black holes in AdS result in logarithmic entropy corrections \cite{izs5}: 
\begin{equation}
S=S_0-\alpha\ln S_0+\cdots, \label{is2}
\end{equation}
where $S_0$ is the typical Bekenstein-Hawking term and $\alpha$ is the correction parameter that differs for various black holes. However, not all of these methods, including Euclidean partition function, canonical ensemble, and AdS/CFT, agree on the computation of such logarithmic corrections (see \cite{izs6} for a review). In this respect, a variety of techniques have been used for entropy computation using microscopic state counting in extreme black holes and other conditions \cite{izs7,izs8,izs9,izs10,izs11,izs12,izs13,izs14,izs15,izs16}. Small thermal fluctuations' impact on the thermodynamics of small black holes including, for example, Schwarzschild–Beltrami–de Sitter black hole, G\"{o}del black hole, Horava–Lifshitz black hole, Van der Waals black hole, AdS black holes etc. have recently become the subject of extensive research (see Refs. \cite{izs17,izs18,izs19,izs20,izs21} and references therein). Here, we investigate how thermal fluctuations affect the thermodynamics of black
holes with constant topological Euler density coupled to Born-Infeld like NLED fields. To this end, we first study the first law of thermodynamics and Smarr's formula. Next, we derive the Hawking temperature, heat capacity, Helmholtz free energy, pressure, enthalpy, and Gibbs free energy \cite{izs22}. We then compute the leading-order logarithmic corrections to the entropy of the black hole caused by thermal variations. We conduct a comparison analysis by plotting a graph between the corrected and non-corrected entropy with regard to the event horizon radius in order to examine the impact of this correction on the entropy. We also compute the internal energy's first-order correction. We plot the corrected and uncorrected internal energy versus the event horizon radius graph to illustrate the change brought on by the thermal fluctuation. We further derive the corrected Helmholtz free energy, corrected pressure, corrected enthalpy, and corrected Gibbs free energy. In the sequel, we do a comparison analysis by graphing the corrected and uncorrected effects on those physical quantities. Finally, we check the stability of those black
holes having constant topological Euler density coupled to Born-Infeld like NLED fields under the thermal fluctuations.}  \\
In the second part, the formation of a naked singularity in the considered spacetime \citep{6}, will be investigated within the framework of quantum mechanics. It has long been known that the Born-Infeld NLED was introduced to eliminate the curvature singularities at the core of charged black holes \citep{7}. However, in the considered model of spacetime \citep{6}, depending on the constant parameters, there exists a classical curvature singularity.  Since the singularities occur at the Planck regime, the analysis of the singularities of general relativity necessitates the use of the tools of quantum mechanics. As an example, a prescription formulated in quantum theory \citep{8,9}, which uses quantum wave packets for probing the singularities states that the unique time evolution of the quantum wave packets for all time implies nonsingular spacetime. Otherwise, the spacetime is said to be quantum singular. This prescription is valid only for static space-times admitting timelike singularities.    \\
The main idea in this formalism is to probe the singularities with quantum wave packets rather than particles. So far, the number of spacetimes including the modified gravity theories has been probed with quantum wave packets to see if the classical singularities are healed or not. For example, the formation of a naked singularity in a model of $\mathcal{F}(R)$ gravity coupled with linear electrodynamics has been considered in \citep{10}. The singularity is probed with quantum wave packets obeying Klein-Gordon, Dirac, and Maxwell fields. The analysis has shown that the classical singularity remains singular. Naked singularities in the weak field regime of $\mathcal{F}(R)$  global monopole spacetime are also studied with spin-$0$, spin-$1/2$, and spin-$1$ quantum wave packets in \citep{11}. The analysis has revealed that the spacetime is quantum singular against scalar field probes, but for the specific modes of Dirac and Maxwell field probes, it becomes quantum regular. The formation of naked singularities in a 5-dimensional Boulware-Deser solution of Einstein-Gauss-Bonnet gravity is considered in \citep{12}.  The quantum singularity analysis of the 5-dimensional Boulware-Deser solution revealed that the classical singularity can be healed when it is probed with scalar fields. Furthermore, the quantum nature of naked singularities developed in the 7-dimensional Lovelock theory is considered in \citep{13}. Again, the classical singularity at $r=0$ is shown to be quantum regular when probed with a scalar field. Moreover, the quantum singularity is studied in a quantum cosmological model within the context of Horava-Lifshitz gravity \citep{14}. It is shown that the quantum Friedmann-Lemaitre-Robertson-Walker universe, which is non-asymptotically flat and filled with radiation in the context of Horava-Lifshitz gravity is quantum mechanically regular. A similar result is obtained in a black hole solution that admits naked singularity in Horava-Lifshitz gravity in \citep{15}. \\
The organization of the paper is as follows. In Sec. \eqref{sec2}, the solution admitting black holes with constant topological Euler density coupled to Born-Infeld like NLED fields is introduced. In addition to black hole solutions, the possibility of having timelike naked singularity is also presented. {\color{black} Section \eqref{sec3} is devoted to the thermodynamical characteristics of those NLED black holes. To describe the thermodynamics of black holes under the influence of thermal fluctuations, we derive a number of additional equations of state in this section. We then derive a generic (corrected) expression for the energies, pressure, enthalpy, and entropy. We also examine the thermal stability of those NLED black holes.} In Sec. \eqref{sec4}, we first give a brief review of the quantum singularity analysis. Then, the naked singularity analyses for two types of quantum waves having different spins [scalar (spin-$0$) and  Dirac (spin-$\frac{1}{2}$) fields] are discussed, respectively. We draw our conclusions in Sec. \eqref{sec5}.

{\color{black}\section{Brief Review of $\mathcal{F(R, G)}$ Gravity and Its Corresponding Black Hole Solution}}   \label{sec2}

{\color{black} $\mathcal{F(R, G)}$ gravity is a modified theory of gravity that extends the standard Einstein-Hilbert action by incorporating additional terms involving the Ricci scalar ($\mathcal{R}$) and the Gauss-Bonnet invariant ($\mathcal{G}$). The motivation behind this modification is to address certain shortcomings of general relativity and explore alternative explanations for the observed accelerated expansion of the universe. This theory has been proposed as a possible candidate for explaining dark energy, dark matter, and other cosmic phenomena that remain unexplained by standard general relativity. Moreover, this theory has garnered substantial attention in the academic community (see for example \cite{isAtazadeh:2013cz,isdelaCruz-Dombriz:2011oii,isElizalde:2020zcb} and references therein) and it is attributed not only to its inherent stability but also to its capacity to effectively elucidate the current acceleration of the universe \cite{isSupernovaSearchTeam:1998fmf}. Additionally, the theory is recognized for its ability to accurately describe phenomena such as the crossing of the phantom divide line \cite{isAhmed:2019inb} and the transition from acceleration to deceleration phases \cite{isGiostri:2012ek}.}

The constant topological Euler density black hole solution, driven by Born-Infeld type Nonlinear Electrodynamics (NLED), was originally presented by Barguena and Vagenas in \citep{6}. In this recently discovered nonlinear electromagnetic solution, the Lagrangian describing the solution has been reformulated utilizing the $P$ framework (formalism) as detailed in \citep{sss1}. Furthermore, it is noteworthy that other significant solutions in NLED have been investigated using the $P$ formalism, as discussed in Ref. \citep{sss2}. The system under examination can be defined and characterized through an auxiliary field denoted by \textcolor{black}{$P_{\mu\nu}=\frac{d\mathcal{L}}{d\mathcal{F}}\mathcal{F}_{\mu\nu}\equiv \mathcal{L}_{\mathcal{F}}\mathcal{F}_{\mu\nu}$ with $\mathcal{F}_{\mu\nu}=E(r)\left[\delta_{\mu}^t\delta_{\nu}^r-\delta_{\mu}^r\delta_{\nu}^t \right]$. Note that, $E(r)$ is the electrical field.} The Legendre transformation corresponding to the dual representation of this system is given by
\begin{equation}
\mathcal{H}=2\mathcal{F}\frac{d\mathcal{L}}{d\mathcal{F}}-\mathcal{L}.\label{sss1}
\end{equation}
Note that the invariant is given by $P=-\frac{1}{4}P_{\mu\nu}P^{\mu\nu}.$Hence, the Lagrangian $\mathcal{L}$, which depends on $P_{\mu\nu}$, can be formulated as
\begin{equation}
\mathcal{L}=2P\frac{d\mathcal{H}}{dP}-\mathcal{H}\label{sss2}
\end{equation}
and the electromagnetic field is
\begin{equation}
F_{\mu\nu}=\frac{d\mathcal{H}}{dP}P_{\mu\nu}.\label{sss3}
\end{equation}
\textcolor{black}{The Hamiltonian-like quantity in the considered model is expressed as equation (A.1) in Ref. \citep{6}. At this stage, we would like to point out that the typos found in the original article (\citep{6}) were corrected through personal conversations with the authors of the article in question. The correct constants should be $\tilde{a}=\frac{4\Lambda^2}{9}$ and $\tilde{b}=\frac{4}{3}q^2\Lambda$, which consequently lead to}
\textcolor{black}{\begin{equation}
\mathcal{H}(P)=\frac{2P-\Lambda}{\sqrt{1-\frac{6P}{\Lambda}}}.\label{sss4}
\end{equation}
With reference to \citep{6}, it is important to state that for small fields $P<<\Lambda$, Eq. \eqref{sss4} simplifies to
\begin{equation}
\mathcal{H}(P)\approx-\Lambda-P+\mathcal{O}(P^2).\label{llll1}
\end{equation}
When the Born-Infeld Hamiltonian is expanded for small fields compared to the maximal field strength $b$, one can get
\begin{equation}
\mathcal{H}_{BI}=-F+\frac{F^2}{2b^2}+\mathcal{O}(F^3). 
\end{equation}
Therefore, the linear Maxwell theory is recovered for very large values of $b$, which in turn implies that in the considered model of NLED this corresponds to when
\begin{equation}
 b^2=\frac{1}{24\sqrt{\Lambda}} .  
\end{equation}
Thereby, the large value of $b$ is possible if $\Lambda$ is negligibly small. 
If we put Eq. \eqref{sss4} into Eq. \eqref{sss2}, the Lagrangian can be rewritten as
\begin{equation}
 \mathcal{L}=\frac{\Lambda(1-\frac{10P}{\Lambda})}{(1-\frac{6P}{\Lambda})^{3/2}}. \label{llll2}  
\end{equation}
For small fields approximation, Eq. \eqref{llll2} reduces to 
\begin{equation}
\mathcal{L}\approx  \Lambda -P+ \mathcal{O}(P^2).
\end{equation}
As a result, a regime corresponding to the linear Maxwell limit occurs when the value of $\Lambda$ is negligibly small. \\                       
The relationship between the $P_{\mu\nu}$ tensor and the electromagnetic potential 1-form $A_\mu = A(r)\delta_\mu^t$ or $A = A(r)dt$ is as follows: First, let us consider the electromagnetic 2-form as
\begin{equation}
F=E(r)dt \wedge dr.   
\end{equation}
Therefore, the electromagnetic field 2-form and electromagnetic dual 2-form are nothing but
\begin{equation}
F=dA=-A(r)'dt\wedge dr \;\; and \;\; ^{*}F=E(r)r^2sin\theta(d\theta\wedge d\varphi),
\end{equation}
where $^*$ represents the dual and $'$ denotes the derivative with respect to the $r$ coordinate. One can easily observe that $E(r)=-A'(r)$. Since $d(^{*}F\mathcal{L}_{\mathcal{F}})=0 \rightarrow Er^2\mathcal{L}_{\mathcal{F}}=constant=-q$. $P_{\mu\nu}$ tensor becomes
\begin{equation}
P_{\mu\nu}= \mathcal{L}_{\mathcal{F}}\mathcal{F}_{\mu\nu}=-\frac{q}{r^2}(\delta_{\mu}^t\delta_{\nu}^r-\delta_{\mu}^r\delta_{\nu}^t).
\end{equation}
Therefore, the electric potential 1- form can be calculated as
\begin{equation}
A(r)=-\int_{const.}^{r} E(r)dr,
\end{equation}
where $E(r)$ is given in Eq. \eqref{m2}. Once $E(r)$ is determined by selecting the appropriate constants, $A(r)$ can subsequently be derived.}

The solution of field equations describes the geometry of a spherically symmetric and static black hole with an endowed metric \citep{6}:
\begin{equation}
ds^{2}=-f(r)dt^{2}+\frac{dr^{2}}{f(r)}+r^{2}d\theta^{2}+r^{2}\sin^{2}\theta d\varphi^{2}. \label{s1}
\end{equation}
The corresponding metric function is given by

\begin{equation}
f(r)=1\pm\sqrt{1-2A+Br+\frac{kr^{4}}{24}}. \label{m1}
\end{equation}

Here, $A$ and $B$ are arbitrary constants, whereas $k$ is the contribution from the topological Euler density defined by

\begin{equation}
\mathcal{G}=R^{\mu\nu\rho\sigma}R_{\mu\nu\rho\sigma}-4R^{\mu\nu}R_{\mu\nu}+R^{2}=k.
\end{equation}

The Einstein-NLED system gives the electric field in terms of these constant parameters as

\begin{equation}
E(r)=\frac{\sqrt{6}}{q}\left\{\frac{4(1-2A)^{2}+48(1-2A)Br+27B^{2}r^{2}+(-1+2A)kr^{4}}{(24-48A+24Br+kr^{4})^{3/2}}\right \}.\label{m2}
\end{equation}

The arbitrary constants $A$ and $B$ can be chosen appropriately so that metric function \eqref{m1}  and the electric field \eqref{m2} asymptotically gives the Reissner-Nordstr\"{o}m de-Sitter solution. While doing so, metric function \eqref{m1} is first expanded asymptotically $(r \rightarrow \infty )$, which yields

\begin{equation}
f(r) = 1 - \sqrt{\frac{6}{k}} \frac{B}{r} - \sqrt{\frac{6}{k}} \frac{2A-1}{r^{2}} - \sqrt{\frac{k}{6}} \frac{r^{2}}{2}.
\end{equation}
In similar fashion, the electric field \eqref{m2}, in the limit as $r \rightarrow \infty $, becomes

\begin{equation}
E(r)=\sqrt{\frac{6}{k}}\left(2A-1 \right).
\end{equation}

If the constants are taken as  $A=\frac{1}{2}+\frac{q^{2}\Lambda}{3}$, $B=\frac{4M\Lambda}{3}$, and $k=\frac{8\Lambda^{2}}{3}$,
then the solution becomes asymptotically Reissner-Nordström de-Sitter and the electric field becomes $ E(r)=\frac{q}{r^{2}}$ as expected.

Substituting these constants into Eq. \eqref{m1}, the solution of the coupled Einstein-NLED system for a certain electromagnetic Hamiltonian given in Appendix of Ref. \cite{6} is obtained as follows
\begin{equation}
f(r)=1\pm\sqrt{\frac{4M\Lambda r}{3}+\frac{\Lambda^{2}r^{4}}{9}-\frac{2q^{2}\Lambda}{3}}.\label{m3}
\end{equation}
Note that the negative sign in Eq. \eqref{m3} implies the black hole solutions. {\color{black} Furthermore, it is noteworthy to mention that, as indicated in \cite{6,isBronnikov:2000vy}, the coupled Einstein-NLED equations are formulated about $P$ and $\mathcal{H}(P)$ in a manner that can be expressed by the equation:
\begin{equation}
\mathcal{H}(P)=-\frac{1}{r^2} \frac{\mathcal{M}(r)}{dr},
\end{equation}
where the mass function $\mathcal{M}(r)$ \cite{isBronnikov:2000vy} determines the metric function given in Eq. \eqref{m3} and it leads to constant (ADM mass: $M$ \cite{isRadinschi:2018nxb}) when $P<<$ (see Ref. \cite{6}).}

The characteristic central singularity at $r=0$ is also encountered in this solution. To this end, the Kretschmann scalar for metric \eqref{s1} is calculated and it is given in terms of the metric function $f(r)$ as

\begin{equation}
\mathcal{K}=R^{\mu\nu\rho\sigma}R_{\mu\nu\rho\sigma}=\left(f^{''}(r)\right)^{2}+\frac{4\left(f^{'}(r)\right)^{2}}{r^{2}}+\frac{4\left(f(r)-1\right)^{2}}{r^{4}},
\end{equation}
which results in

\begin{equation}
\mathcal{K}=\frac{36b^{2}r^{4}-3br^{2}\left( a+4br^{3} \right)^{2}}{\left( ar+br^{4}-c \right)^{2}}+\frac{\left( a+4br^{3} \right)^{4}}{16\left( ar+br^{4}-c \right)^{3}}+\frac{\left( a+4br^{3} \right)^{2}}{r^{2}\left( ar+br^{4}-c \right)}+\frac{4\left( ar+br^{4}-c \right)}{r^{4}},
\end{equation}
where $a=\frac{4M\Lambda }{3}$, $b=\frac{\Lambda^{2}}{9}$, and $c=\frac{2q^{2}\Lambda}{3}$.
The diverging character of the curvature invariant $\mathcal{K}$ is evident when $r \rightarrow 0$. \\
Bargueno and Vagenas have studied the number of massive and massless black hole solutions. However, there is a class of interesting massive black hole solutions, which was not mentioned in their analysis. This solution is obtained if the cosmological constant-related parameter is taken as $ \Lambda = - \frac{3}{2q^{2}} $. The metric function \eqref{m3} for this particular value of $ \Lambda $ becomes,
\begin{equation}
  f(r)=1-\sqrt{\frac{r^{4}}{4q^{4}}-\frac{2Mr}{q^{2}}+1}. \label{m4}
\end{equation}
This particular class of black hole has a single event horizon obtained from $f(r_{h})=0$, which gives $r_{h}=(8Mq^{2})^{1/3}$. As was stated by Barguena and Vagenas, the central singularity at $r=0$ becomes timelike naked whenever the positive sign in Eq. \eqref{m3} is taken or if we consider massless solution $(M=0)$ with $ \Lambda = - \frac{3}{2q^{2}} $. \\
Our aim in this paper is two-fold. First, we analyze the thermodynamic characteristics of the massive black hole described with the metric function \eqref{m4}. As a second aim, we investigate the appearance of timelike naked singularities in view of quantum mechanics.
 
\section{ Thermodynamic Characteristics of Black Hole Solutions } \label{sec3}
This section is devoted to analyzing the thermodynamic characteristics of the particular class of topological black hole solution described in Eqs. \eqref{is1} and \eqref{m4} obtained in the presence of NLED in Ref. \cite{6}. Thermodynamic quantities such as Hawking temperature, Helmholtz free energy, internal energy, volume, entropy, Gibbs free energy, and leading order corrections to thermodynamic potentials are derived. \\

\subsection{The First Law of Thermodynamics and Smarr's Formula}

The homogeneous function theorem of Euler is used to create the Smarr's formula, which states that if $f(\lambda^{i}x,\lambda^{j}y,\lambda^{k}z)=\lambda^{l}f(x,y,z)$, in which $\lambda$ denotes  a  constant  and $(i,j,k)$ are nothing but the integer powers \cite{m}, then:
\begin{equation}
f(x,y,z)=l^{-1} \bigg[i\frac{\partial f}{\partial x}x+j\frac{\partial f}{\partial y}y+k\frac{\partial f}{\partial z}z\bigg].
\end{equation}

Firstly, let us calculate the mass of the black hole by using the metric function defined in Eq. \eqref{m4} to be zero at the point $r=r_{H}$.
The  mass of  the  black  hole  can  be  written  as

\begin{equation}
M(q,\Lambda,r_{H})=\frac{q^{2}}{2r_{H}}-\frac{\Lambda r^{3}_{H}}{12}-\frac{3}{4\Lambda r_{H}}.\label{m5}
\end{equation}

The Hawking-Bekenstein entropy of four-dimensional spherical symmetric black holes is given by (in geometric units: $G=c=\hbar=1$).

\begin{equation}
S=\pi r^{2}_{H}.\label{m6}
\end{equation}

If  we  use  Eq. \eqref{m6} in  mass  function  \eqref{m5}, we can express the mass function in terms of the entropy:

\begin{equation}
M(S,q,\Lambda)=\frac{q^{2}\pi^{1/2}}{2S^{1/2}}-\frac{\Lambda S^{3/2}}{12\pi^{3/2}}-\frac{3\pi^{1/2}}{4S^{1/2}\Lambda }.\label{m7}
\end{equation}

Now we  reformulate  the  variables  of    mass  function \eqref{m7} as demonstrated
\begin{equation}
S \rightarrow \lambda^{i} S, \;\;\; \;  q \rightarrow \lambda^{j} q\;\;\; \; and  \;\;\; \;  \Lambda \rightarrow \lambda^{k} \Lambda .
\end{equation}

Moreover, the  integer  powers  of  reformulated  parameters can satisfy Euler's  homogeneous  function  theorem as follows

\begin{equation}
i=2l, \;\;\; \;  j=l,  \;\;\; \;  k=-2l ,
\end{equation}

in light of this information, the Smarr's  formula can be reduced to

\begin{equation}
M(S,q,\Lambda)=2S\left(\frac{\partial M}{\partial S}\right)+q\left(\frac{\partial M}{\partial q}\right)-2\Lambda\left(\frac{\partial M}{\partial \Lambda}\right).\label{m9}
\end{equation}

The  derivative  of Eq.\eqref{m7}  with  respect  to  the  entropy  gives   the  Hawking  temperature,

\begin{equation}
T_{H}=\frac{f^{`}(r_{H})}{4\pi}=\frac{\partial M}{\partial S}.\label{m8}
\end{equation}

When  we  employ Eq. \eqref{m8} in the computation of the  Hawking  temperature ($T_{H}=\frac{\kappa}{2\pi}$) \cite{iswald,isrev}, one can obtain the following expression

\begin{equation}
T_{H}=\frac{\partial M}{\partial S}=\frac{3\pi^{1/2}}{8S^{3/2}\Lambda}-\frac{q^{2}\pi^{1/2}}{4S^{3/2}}-\frac{S^{1/2}\Lambda}{8\pi^{3/2}}.\label{m10}
\end{equation}

The second term in  Eq.\eqref{m9}  is the electric potential energy on the horizon.  Thus, the electric potential ($\Phi_{q}$ ) is  given  by

\begin{equation}
\Phi_{q}=\frac{\partial M}{\partial q}=\sqrt{\frac{\pi}{S}}q.
\end{equation}

The  derivative  of  the  black  hole  mass  with  respect  to  $\Lambda$  is

\begin{equation}
V=\frac{\partial M}{\partial \Lambda}=\frac{3\pi^{1/2}}{8S^{3/2}\Lambda^{2}}-\frac{1}{12}\left( \frac{S}{\pi}  \right)^{3/2}.
\end{equation}

Then, the Smarr's formula can be written as

\begin{equation}
M=2ST_{H}+q\Phi_{q}-2\Lambda V.
\end{equation}

For the analysis of the heat capacity and thermodynamic potentials, we insert  Eq.\eqref{m6}  in   Eq.\eqref{m10}. Thus, Eq.\eqref{m10} transforms into

\begin{equation}
T_{H}=\frac{3-2q^{2}\Lambda-\Lambda^{2}r^{4}_{H}}{8\pi \Lambda r^{3}_{H}}. \label{s23}
\end{equation}

 If the cosmological constant related parameter is taken as $ \Lambda = - \frac{3}{2q^{2}} $,  Eq. \eqref{s23} reduces to

\begin{equation}
T_{H}=\frac{3r^{4}_{H}-8q^{4}}{16\pi q^{2} r^{3}_{H}}. \label{s24}
\end{equation}

\begin{figure}[h]
  \centering
  \includegraphics[scale=0.6]{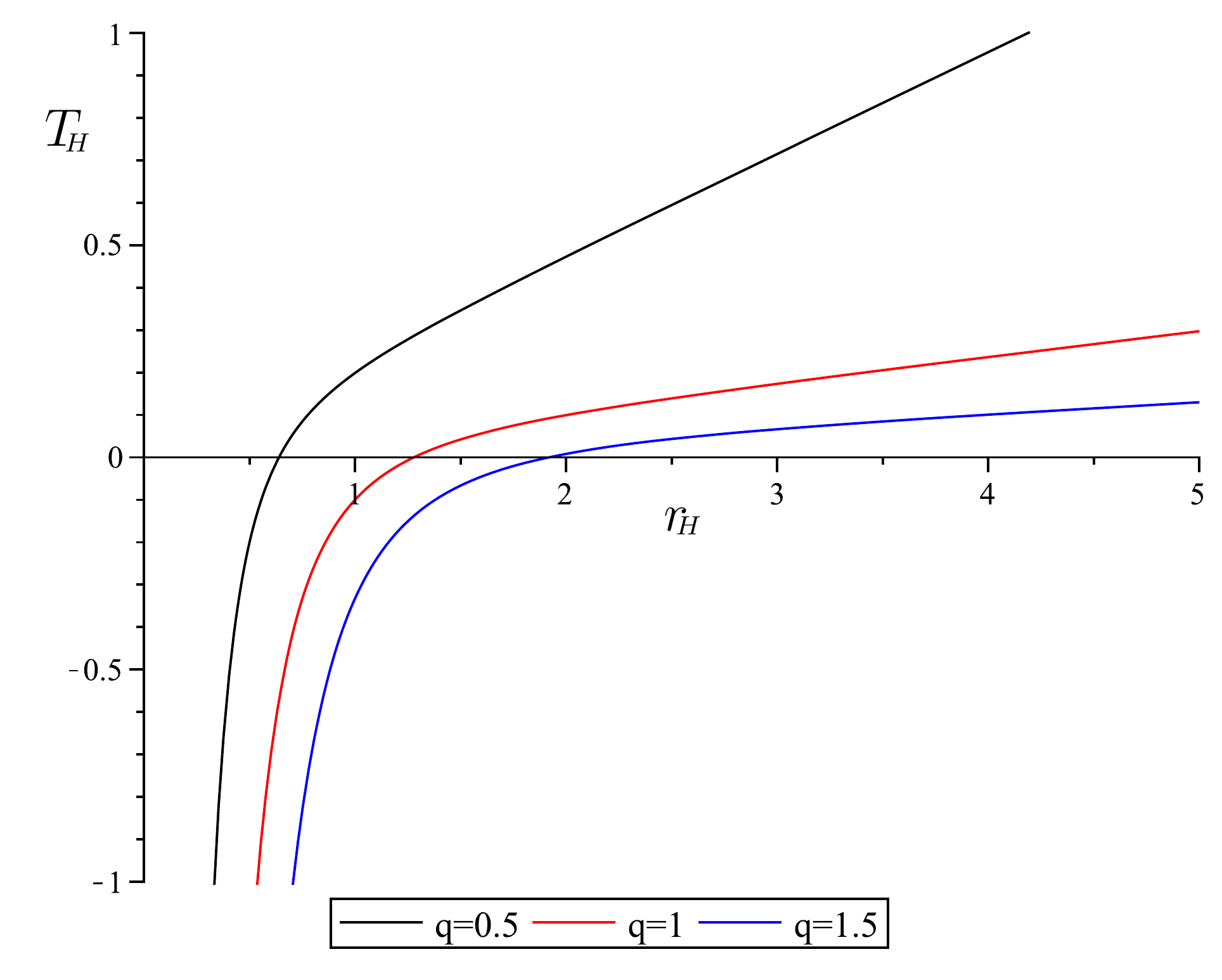} 
  \caption{Hawking temperature versus the horizon radius for various $q$ values.} \label{fig1}
\end{figure}

 \newpage

{\color{black}The behavior of the Hawking temperature versus the horizon radius can be seen from
Fig. \eqref{fig1}. Here, one can see that the Hawking temperature shows an increasing behavior with increasing
 event horizon radius and decreasing $q$ value. There exists a zero temperature case in which the system must be in a state with the least amount of energy. There is often just one state (known as the ground state) with the least amount of energy, and entropy is proportional to the number of accessible microstates. The entropy at absolute zero will then be precisely zero in such a scenario.}

The heat  capacity  is given  by

\begin{equation}
C_{q}=T_{H}\left(\frac{\partial S}{\partial T_{H}}\right), \label{s25}
\end{equation}

we  obtain

\begin{equation}
C_{q}=\frac{2 \pi}{3}\left( \frac{3 r^{6}_{H}-8 r^{2}_{H}q^{4}}{8q^{4}+ r^{4}_{H}}   \right). \label{s26}
\end{equation}

The  Helmholtz  free  energy  (denoted  by  $F$)    of  the  black  hole  is given by

\begin{equation}
F=-\int SdT_{H}=-\int \pi r^{2}_{H} \left[ \frac{3}{16\pi q^{2}}+\frac{3q^{2}}{2\pi r^{4}_{H}} \right]  dr_{H}, \label{s27}
\end{equation}

and  the  solution  of  integral  \eqref{s27} reads

\begin{equation}
F=\frac{3q^{2}}{2r_{H} }-\frac{r^{3}_{H}}{16 q^{2}}.
\end{equation}

As is well-known, the  internal  energy  ($E$)  of  the  system is given by

\begin{equation}
E=\int T_{H} dS. \label{s29}
\end{equation}

When we write the temperature \eqref{s23}  and entropy \eqref{m6} into Eq. \eqref{s29}, the internal energy becomes

\begin{equation}
E=\frac{r^{3}_{H}}{8 q^{2}}+\frac{q^{2}}{r_{H} }.
\end{equation}

Also,  the  volume  of  the black  hole can be calculated  as

\begin{equation}
V=4\int S dr_{H}=4\int \pi r^{2}_{H} dr_{H}=\frac{4}{3}\pi r^{3}_{H}.
\end{equation}

Furthermore, the following derivatives to be used for calculating the pressure  ($P$) of the black hole are

\begin{equation}
\frac{dV}{dr_{H}}=4\pi r^{2}_{H}
\end{equation}

and  

\begin{equation}
\frac{dF}{dr_{H}}=-\frac{3q^{2}}{2r^{2}_{H}}-\frac{3r^{2}_{H}}{16q^{2} }.
\end{equation}

Recalling the definition of Helmholtz free energy $(F=E-TS)$, one can compute the pressure of the black hole as follows 

\begin{equation}
P=-\frac{dF}{dV}=-\frac{dF}{dr_{H}}\frac{dr_{H}}{dV}=\frac{3q^{2}}{8\pi r^{4}_{H}}+\frac{3}{64\pi q^{2} }.
\end{equation}

Besides, the enthalpy ($H$) \cite{n17} is defined by

\begin{equation}
H=E+PV. \label{s35}
\end{equation}

Substituting the obtained values of internal energy, pressure, and volume in Eq.\eqref{s35}, we find

\begin{equation}
H=\frac{3r^{3}_{H}}{16 q^{2}}+\frac{3q^{2}}{2r_{H} }. \label{s36}
\end{equation}

The thermodynamical formula for Gibbs free energy ($G$) is given by

\begin{equation}
G=F+PV. \label{s37}
\end{equation}

If we put the values of $F$, $P$ and $V$ into Eq.\eqref{s37} , the Gibbs free energy becomes

\begin{equation}
G=\frac{2q^{2}}{r_{H} }.
\end{equation}

\subsection{Logarithmic Corrections to Black Hole Entropy}

\textcolor{black}{The system in thermal equilibrium has the following density of states \cite{nn17,n17}
\begin{equation}
\rho(E)=\frac{e^{S_{0}}}{\sqrt{2\pi \frac{d^2S}{d\beta^2} }},    
\end{equation}
where $S_0$ is the equilibrium value of the entropy and $\beta=1/T_H$. Since the logarithm of the density of states defines the entropy of the microcanonical ensemble, the entropy is expressed as
\begin{equation}
  S\equiv ln\rho=S_{0}-\frac{1}{2} ln CT^{2}_{H}+ (sub \;\; leading \;\; terms).
\end{equation}
Without loss generality, we introduce a new variable ($\alpha$) into the second term to parameterize the role of thermal fluctuations on the thermodynamics of the system. In this context, the generic logarithmic corrected entropy is given by}

\begin{equation}
S=S_{0}-\frac{\alpha}{2} ln CT^{2}_{H}+(sub \;\; leading \;\; terms),\label{s39}
\end{equation} 

where $S_{0}=\pi r^{2}_{H}$, $\alpha$ and $C$  are the thermal fluctuations parameter and the heat capacity of the systems, respectively \cite{n17}. {\color{black}For maximum thermal fluctuations effects, the correction parameter $\alpha$ in Eq. \eqref{is2} was set to one: $\alpha=1$ (see, for instance, Ref. \cite{n17})}. Substituting the values of Hawking temperature \eqref{s24} and heat  capacity  \eqref{s26}  in Eq. \eqref{s39}, we calculate the microcanonical entropy of the system as follows

\begin{equation}
S=S_{0}-\frac{1}{2}ln\left(\frac{\left( 3r^{4}-8q^{4} \right)^{3}}{128q^{4}r^{4}_{H}\left( 24q^{4}+3r^{4}_{H} \right)}\right). \label{s40}
\end{equation}

\begin{figure}[h]
  \centering
  \includegraphics[scale=0.6]{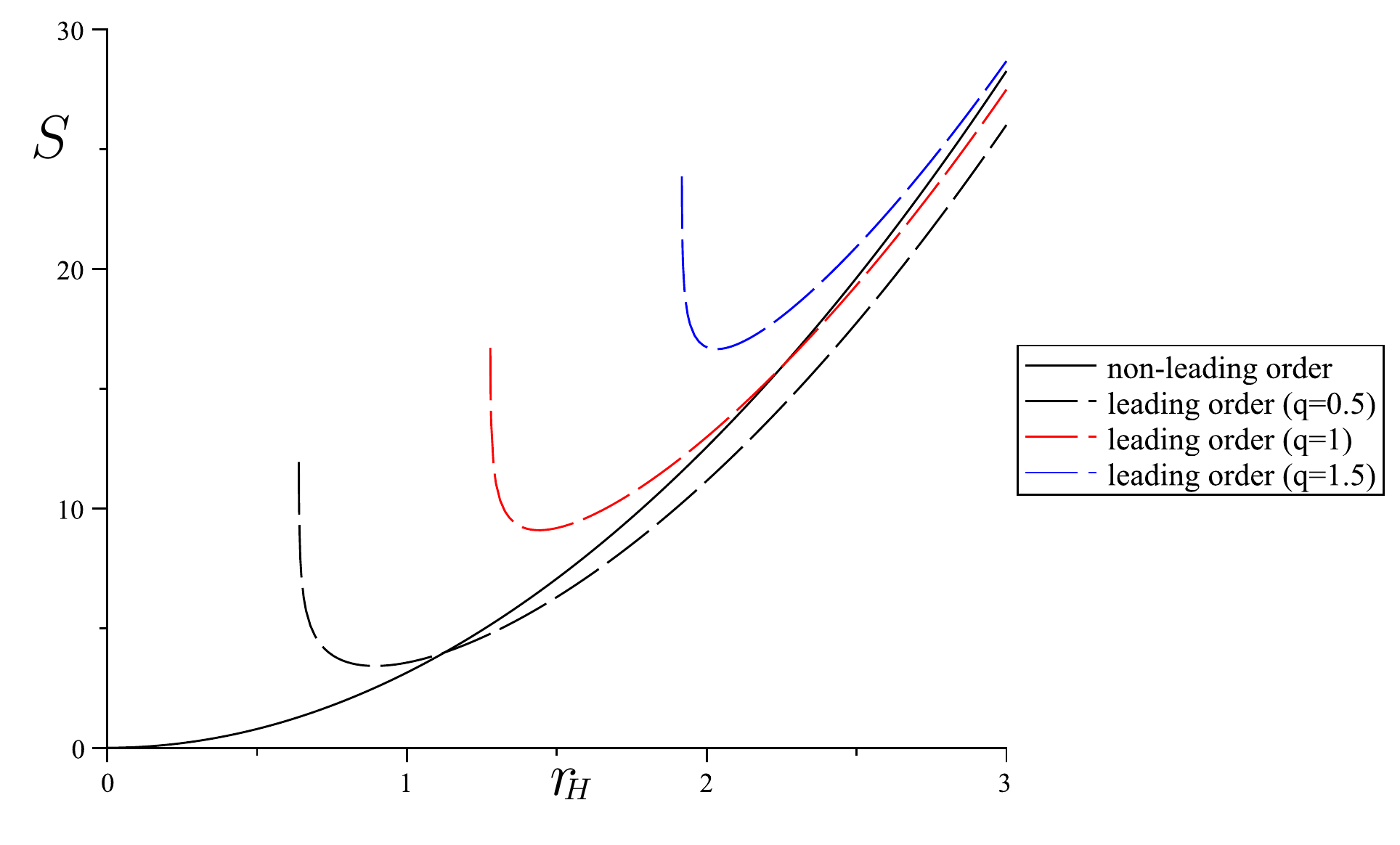}
  \caption{Entropy versus the black hole horizon for various $q$ values. } \label{fig2}
\end{figure} 
\newpage

Figure \eqref{fig2} makes clear that the system has a positive entropy. Furthermore, the corrected entropy behaves as a decreasing function below a crucial horizon radius. But, for the black holes with horizon radii greater than the critical value, the corrected entropy behaves similarly to the equilibrium entropy behavior: it increases as the horizon radius increases. This raises the crucial point in a way that contrary to expectations, tiny thermal variations do not have a significant impact on the thermodynamics of large-sized ($r_{H}>>$) black holes. \textcolor{black}{The other interesting trend is the changes in parabolic entropy due to corrections, and each new entropy having a minimum, which increases for different charge values.}

The corrected  internal  energy  is  given  by

\begin{equation}
E=\int T_{H} dS. \label{s41}
\end{equation}

Writing the resulting values for Hawking temperature \eqref{s24},  entropy \eqref{s40}, and $S_{0}$  in Eq. \eqref{s41}, we obtain

\begin{equation}
\begin{aligned}
E=& \frac{r^{3}_{H}}{8 q^{2}}+\frac{q^{2}}{r_{H} }-\frac{3r_{H}}{8\pi q^{2}}+\frac{q^{2}}{3\pi r^{3}_{H}}-\frac{2^{1/4}}{8\pi q}ln\left[ \frac{r^{2}_{H}+2^{5/2}qr_{H}+2\sqrt{2}}{r^{2}_{H}-2^{5/2}qr_{H}+2\sqrt{2}} \right]\\
& -\frac{2^{1/4}}{4\pi q}\left[tan^{-1}\left( \frac{r_{H}}{2^{1/4}q}+1 \right) + tan^{-1}\left( \frac{r_{H}}{2^{1/4}q}-1 \right) \right].
\end{aligned} \label{s42}
\end{equation}

\begin{figure}[h]
  \centering
  \includegraphics[scale=0.6]{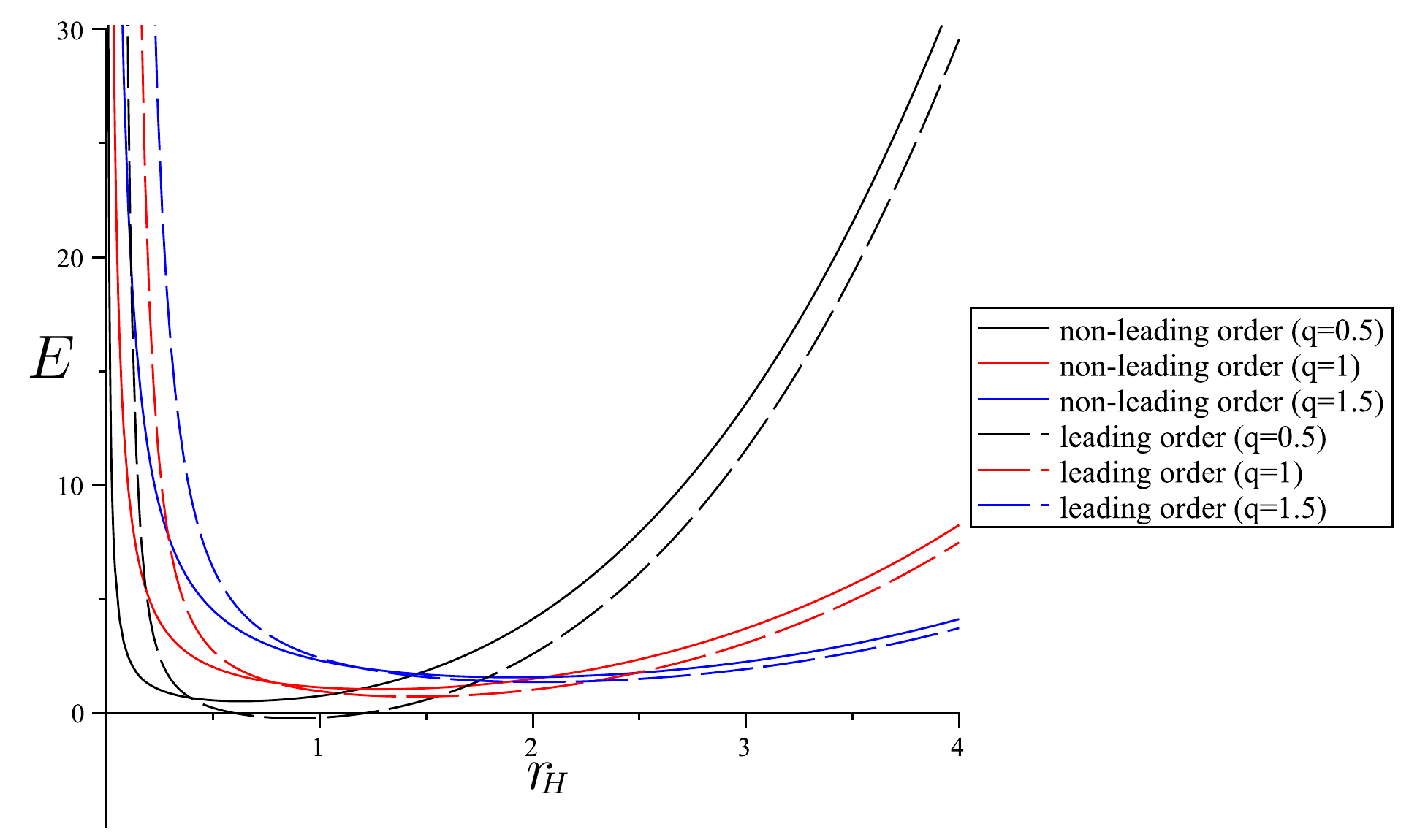}
  \caption{Corrected internal energy versus the black hole horizon for various $q$ values.} \label{fig3}
\end{figure}

The internal energy \eqref{s42} of the system is positive, as can be shown in Fig. \eqref{fig3}. There is a critical horizon radius that the corrected internal energy gets its lowest value. However, for black holes with horizon radii greater than the critical value, the corrected internal energy behavior changes to an increasing function, much like the corrected entropy behavior. \textcolor{black}{ In other words, within the region where $r_H<<$, corrections dominate the internal energy, while within the region where $r_H>>$, corrections do not dominate. Additionally, increases in charge values increase internal energy in the $r_H<<$ region but decrease it in the $r_H>>$ region.}

\newpage
The corrected Helmholtz  free  energy ($F$) is defined by

\begin{equation}
F=-\int SdT_{H}. \label{s43}
\end{equation}

Moreover, if we  put Hawking temperature \eqref{s24}, entropy \eqref{s40}, and $S_{0}$  in Eq. \eqref{s43}, we find

\begin{equation}
\begin{aligned}
F=&\frac{3q^{2}}{2r_{H} }-\frac{r^{3}_{H}}{16 q^{2}}+\frac{1}{4\pi} ln\left(\frac{(3r^{4}_{H}-8q^{2})^{3}}{q^{4}r^{4}_{H}(8q^{4}+4r^{4}_{H})}  \right)\left[ \frac{3r_{H}}{8q^{2}}-\frac{q}{r^{3}_{H}} \right]-\frac{3r_{H}}{8\pi q^{2}}+\frac{q^{2}}{3\pi r^{3}_{H}} \\
&-\frac{2^{1/4}}{8\pi q}ln\left[ \frac{r^{2}_{H}+2^{5/2}qr_{H}+2\sqrt{2}}{r^{2}_{H}-2^{5/2}qr_{H}+2\sqrt{2}} \right]+\frac{q^{2}ln(384)}{4\pi r^{3}_{H}}-\frac{3ln(384)r_{H}}{32\pi q^{2}}\\
&-\frac{2^{1/4}}{4\pi q}\left[tan^{-1}\left( \frac{r_{H}}{2^{1/4}q}+1 \right) + tan^{-1}\left( \frac{r_{H}}{2^{1/4}q}-1 \right) \right].
\end{aligned}
\end{equation}

\begin{figure}[h]
  \centering
  \includegraphics[scale=0.6]{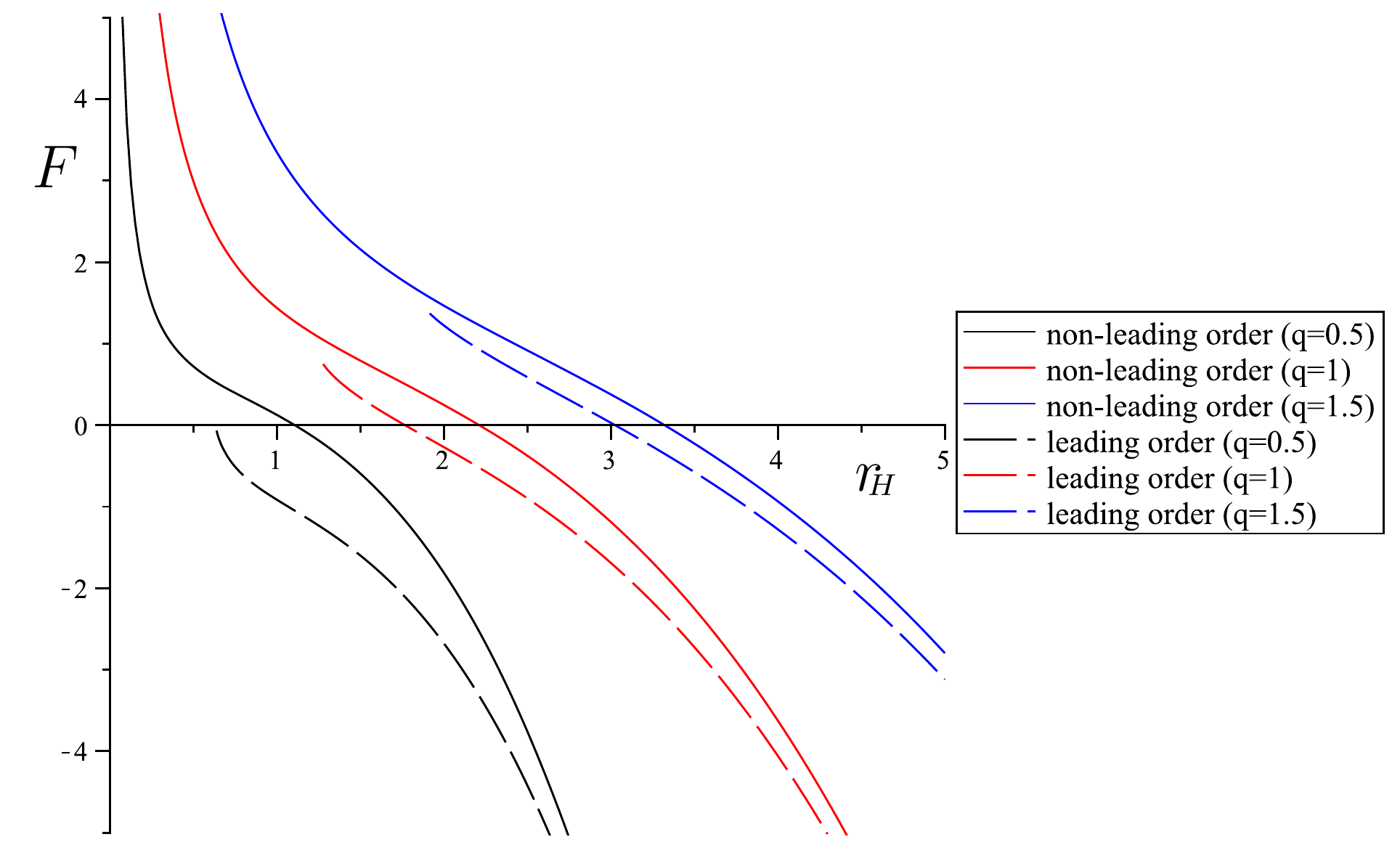}
  \caption{Corrected Helmholtz free energy versus the black hole horizon for various $q$ values.} \label{fig4}
\end{figure}
\newpage

{\color{black} Figure \eqref{fig4} illustrates the Helmholtz free energy's behavior with respect to the event horizon radius. It can be observed from the figure that the Helmholtz free energy behaves in a way that decreases as the radius increases. There is a critical horizon where the free energy is unaffected by thermal variations. The corrected free energy becomes much negative for larger black holes (with a horizon radius larger than the critical horizon). The increasing $q$ parameters, however, make the free energy more positive for smaller black holes.}\\
The  corrected  pressure and free energy  relation  is  given  by

\begin{equation}
P=-\frac{dF}{dV},
\end{equation}

which results in

\begin{equation}
\begin{aligned}
P=&\frac{3q^{2}}{8\pi r^{4}_{H}}+\frac{3}{64\pi q^{2} }+\frac{3q^{2}ln(384)}{16\pi^{2} r^{6}_{H}}+\frac{3ln(384)}{128\pi^{2} q^{2}r^{2}_{H}}+\frac{3}{32\pi^{2}q^{2}r^{2}_{H}}+\frac{q^{2}}{\pi^{2}r^{6}_{H}}\\
&-\frac{3}{64\pi^{2}r^{2}_{H}}ln\left( \frac{(3r^{4}_{H}-8q^{4})^{3}}{q^{4}r^{4}_{H}(8q^{4}+r^{4}_{H})} \right)\left[ \frac{q^{2}}{r^{4}_{H}}+\frac{1}{8q^{2}} \right]\\
&+\frac{1}{16\pi^{2}q^{2}r^{2}_{H}}\left[\frac{1}{1+\left( \frac{r_{H}}{2^{1/4}q}+1 \right)^{2}}  +\frac{1}{1+\left( \frac{r_{H}}{2^{1/4}q}-1 \right)^{2}} \right]\\
&+\frac{2^{1/4}}{32\pi^{2}q^{2}r^{2}_{H}}\left[ \frac{2r_{H}+2^{5/2}q}{r^{2}_{H}+2^{5/2}qr_{H}+2\sqrt{2}}+\frac{2r_{H}-2^{5/2}q}{r^{2}_{H}-2^{5/2}qr_{H}+2\sqrt{2}}    \right]\\
&+\left[\frac{9}{q^{4}r_{H}(3r^{4}_{H}-8q^{2})}-\frac{1}{q^{2}r^{5}_{H}}-\frac{1}{q^{2}r_{H}4(3r^{4}_{H}+8q^{4})}\right]\left(  \frac{q^{4}}{4\pi^{2}r_{H}}-\frac{3r^{3}_{H}}{32\pi^{2}}   \right).
\end{aligned}
\end{equation}

\begin{figure}[h]
  \centering
  \includegraphics[scale=0.6]{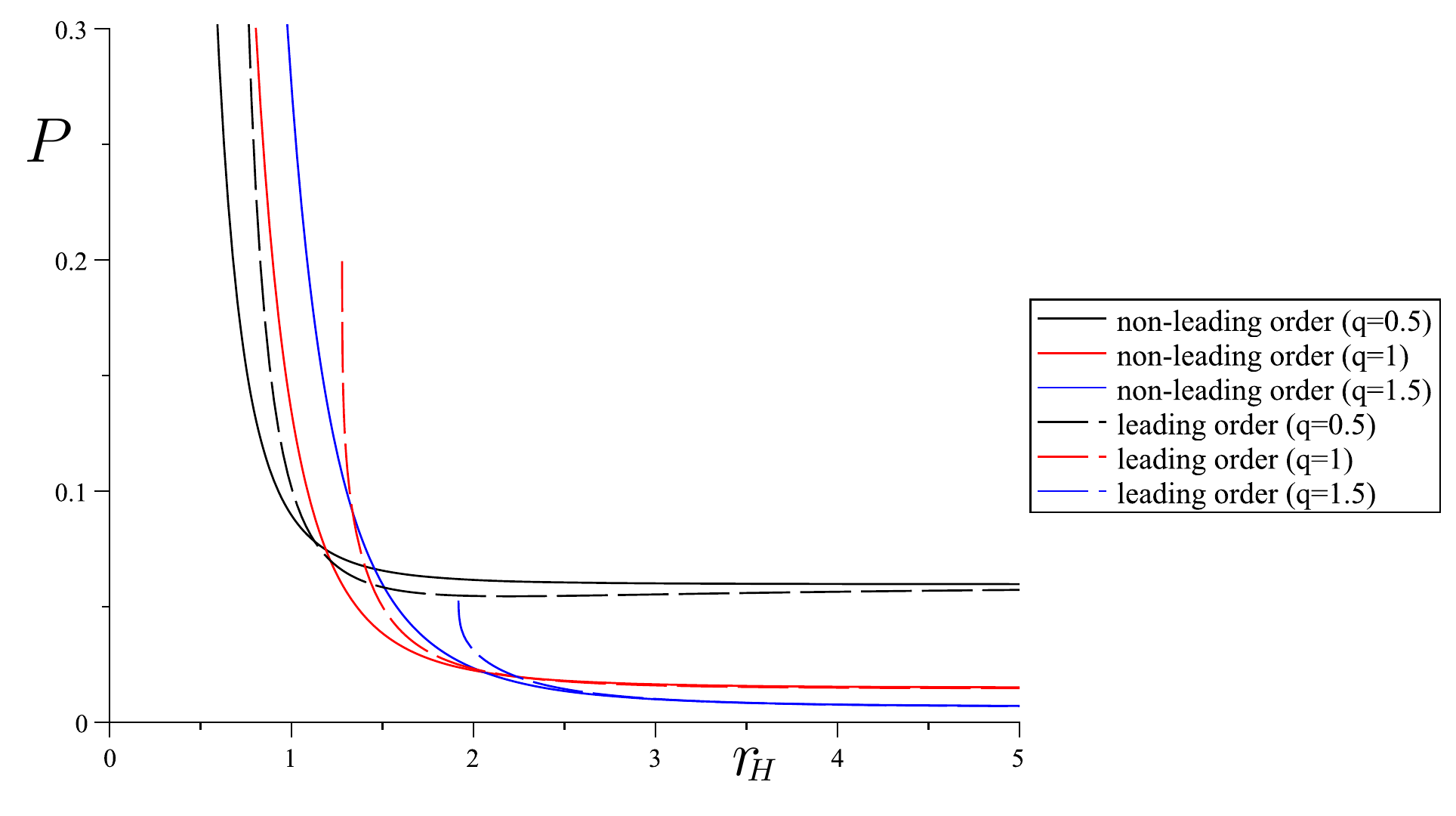}
  \caption{Corrected pressure-free energy versus the black hole horizon for various $q$ values.} \label{fig5}
\end{figure}

{\color{black} It is clear from Fig. \eqref{fig5} that the corrected pressure becomes saturated for the NLED black holes having large values of $r_{H}$. This indicates that for large black holes, the pressure is not much impacted by modest temperature fluctuations. However, due to thermal fluctuations, the pressure rises to a high value for very small black holes (i.e. for very small $r_{H}$). Thus, astonishingly, it is impossible to ignore the contributions of thermal fluctuations to the pressure of the small NLED black holes. In reality, when the horizon radius approaches zero, the adjusted pressure approaches asymptotic values.}

If we use the obtained corrected  pressure, the corrected  internal  energy,  and  the  volume  in Eq. \eqref{s35}, the corrected  enthalpy ($H$) can be calculated as

\begin{equation}
\begin{aligned}
H=& \frac{3r^{3}_{H}}{16 q^{2}}+\frac{3q^{2}}{2r_{H} }-\frac{3r_{H}}{8\pi q^{2}}+\frac{q^{2}}{3\pi r^{3}_{H}}-\frac{2^{1/4}}{8\pi q}ln\left[ \frac{r^{2}_{H}+2^{5/2}qr_{H}+2\sqrt{2}}{r^{2}_{H}-2^{5/2}qr_{H}+2\sqrt{2}} \right]\\
& -\frac{2^{1/4}}{4\pi q}\left[tan^{-1}\left( \frac{r_{H}}{2^{1/4}q}+1 \right) + tan^{-1}\left( \frac{r_{H}}{2^{1/4}q}-1 \right) \right]\\
&+\frac{q^{2}ln(384)}{4\pi r^{3}_{H}}+\frac{ln(384)r_{H}}{32\pi q^{2}}+\frac{r_{H}}{8\pi q^{2}}+\frac{4q^{2}}{3\pi r^{3}_{H}}\\
&-\frac{r_{H}}{16\pi q^{2}}ln\left( \frac{(3r^{4}_{H}-8q^{4})^{3}}{q^{4}r^{4}_{H}(8q^{4}+r^{4}_{H})} \right)\left[ \frac{q^{2}}{r^{4}_{H}}+\frac{1}{8q^{2}} \right]\\
&+\frac{r_{H}}{12\pi q^{2}}\left[\frac{1}{1+\left( \frac{r_{H}}{2^{1/4}q}+1 \right)^{2}}  +\frac{1}{1+\left( \frac{r_{H}}{2^{1/4}q}-1 \right)^{2}} \right]\\
&+\frac{2^{1/4}r_{H}}{24\pi q^{2}}\left[ \frac{2r_{H}+2^{5/2}q}{r^{2}_{H}+2^{5/2}qr_{H}+2\sqrt{2}}+\frac{2r_{H}-2^{5/2}q}{r^{2}_{H}-2^{5/2}qr_{H}+2\sqrt{2}}    \right]\\
&+\left[\frac{9}{q^{4}(3r^{4}_{H}-8q^{2})}-\frac{1}{q^{2}r^{4}_{H}}-\frac{1}{4q^{2}(3r^{4}_{H}+8q^{4})}\right]\left(  \frac{q^{4}r_{H}}{3\pi }-\frac{r^{5}_{H}}{8\pi}   \right).
\end{aligned}
\end{equation}

\begin{figure}[h]
  \centering
  \includegraphics[scale=0.6]{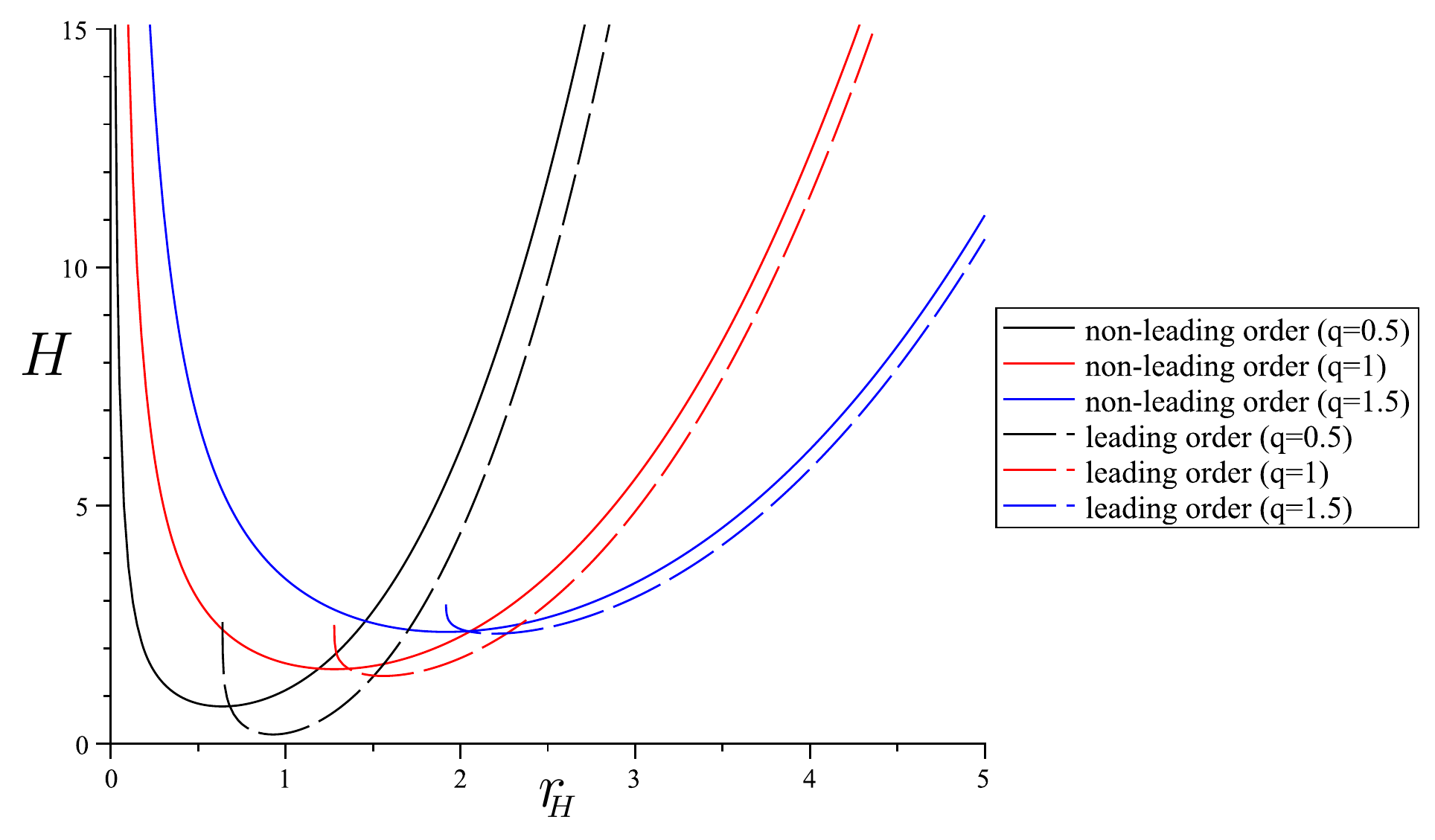}
  \caption{Corrected enthalpy versus the black hole horizon for various $q$ values.} \label{fig6n}
\end{figure}

We plot the graph depicted in Fig. \eqref{fig6n} in order to investigate how thermal fluctuations affect the enthalpy. According to the plot, the enthalpy is always positive and there exists a critical horizon radius that the corrected enthalpy gets its lowest value. The corrected enthalpy behavior, like the corrected energy behavior, switches to an increasing function for the NLED black holes with horizon radii bigger than the critical value. The value of the enthalpy falls as the $q$ parameter value rises. \textcolor{black}{ Additionally, a parabolic trend is observed, similar to the entropy.}

Similarly, when we put  the corrected  pressure, the corrected  free  energy,  and  the  volume  in Eq. \eqref{s37}, the corrected  Gibss free energy ($G$) is found to be

\begin{equation}
\begin{aligned}
G=&\frac{2q^{2}}{r_{H} }-\frac{r^{3}_{H}}{16 q^{2}}+\frac{3q^{2}}{2r_{H}}-\frac{r^{3}_{H}}{16q^{2}}-\frac{3r_{H}}{8\pi q^{2}}+\frac{q^{2}}{3\pi r^{3}_{H}}+\frac{q^{2}ln(384)}{4\pi r^{3}_{H}}-\frac{3ln(384)r_{H}}{32\pi q^{2}}\\
&-\frac{2^{1/4}}{8\pi q}ln\left[ \frac{r^{2}_{H}+2^{5/2}qr_{H}+2\sqrt{2}}{r^{2}_{H}-2^{5/2}qr_{H}+2\sqrt{2}} \right]-\frac{2^{1/4}}{4\pi q}\left[tan^{-1}\left( \frac{r_{H}}{2^{1/4}q}+1 \right) + tan^{-1}\left( \frac{r_{H}}{2^{1/4}q}-1 \right) \right]\\
&+\frac{1}{16q^{2}r^{3}_{H} }+\frac{q^{2}ln(384)}{4\pi r^{3}_{H}}+\frac{ln(384)r_{H}}{32\pi q^{2}}+\frac{r_{H}}{8\pi q^{2}}+\frac{4q^{2}}{3\pi r^{3}_{H}}\\
&-\frac{r_{H}}{16\pi q^{2}}ln\left( \frac{(3r^{4}_{H}-8q^{4})^{3}}{q^{4}r^{4}_{H}(8q^{4}+r^{4}_{H})} \right)\left[ \frac{q^{2}}{r^{4}_{H}}+\frac{1}{8q^{2}} \right]\\
&+\frac{r_{H}}{12\pi q^{2}}\left[\frac{1}{1+\left( \frac{r_{H}}{2^{1/4}q}+1 \right)^{2}}  +\frac{1}{1+\left( \frac{r_{H}}{2^{1/4}q}-1 \right)^{2}} \right]\\
&+\frac{2^{1/4}r_{H}}{24\pi q^{2}}\left[ \frac{2r_{H}+2^{5/2}q}{r^{2}_{H}+2^{5/2}qr_{H}+2\sqrt{2}}+\frac{2r_{H}-2^{5/2}q}{r^{2}_{H}-2^{5/2}qr_{H}+2\sqrt{2}}    \right]\\
&+\left[\frac{9}{q^{4}(3r^{4}_{H}-8q^{2})}-\frac{1}{q^{2}r^{4}_{H}}-\frac{1}{4q^{2}(3r^{4}_{H}+8q^{4})}\right]\left(  \frac{q^{4}r_{H}}{3\pi }-\frac{r^{5}_{H}}{8\pi}   \right).
\end{aligned}
\end{equation}

\begin{figure}[h]
  \centering
  \includegraphics[scale=0.6]{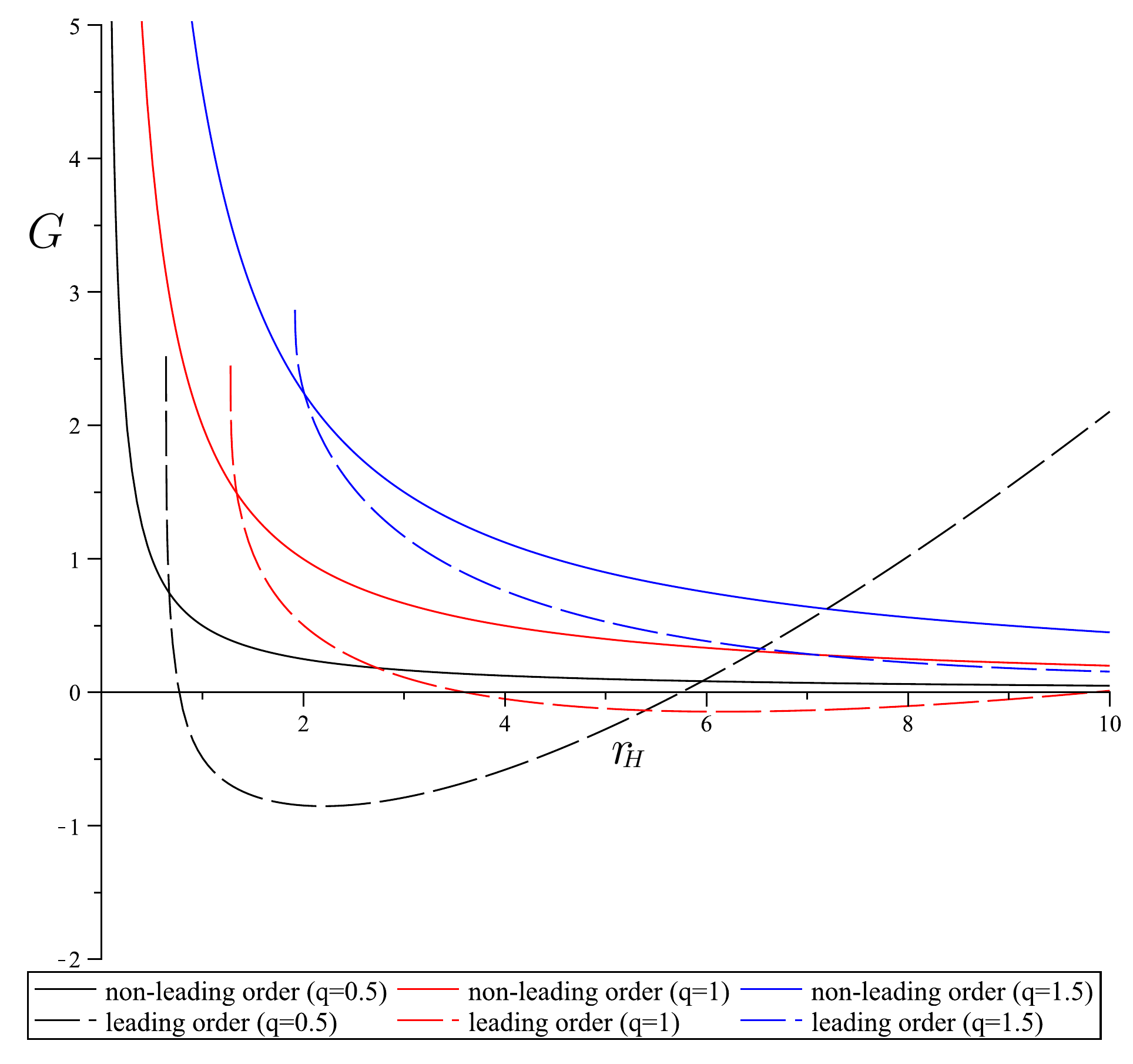}
  \caption{Corrected Gibbs free energy versus the black hole horizon for various $q$ values.} \label{fig7}
\end{figure}
{\color{black} }

\newpage

To study the behavior of the Gibbs free energy and the effects of thermal fluctuations, the graph shown in Fig. \eqref{fig7} is plotted. From the figure, one can see that the Gibbs free energy can take negative values when the small statistical fluctuations are considered. We notice that for small $q$ values (like $q=0.5$), there might be even two critical points and the Gibbs free energy becomes negative among those points \textcolor{black}{and this critical transition charge value decreases and even disappears as it increases.}. Before and after the critical values, $G$ gets positive values again. Furthermore, as the black hole size increases, the asymptotic values of the Gibss free energy are close to each other.

\subsection{Stability}

 The corrected specific heat $(C)$ can be defined with the  help of the following formula 

\begin{equation}
C=\frac{dE}{dT}
\end{equation}

Using Eq.\eqref{s42}, one can thus calculate the expression for leading-order corrected specific heat for the black holes. As a result, leading-order corrected specific heat can be calculated as

\begin{equation}
\begin{aligned}
C=&\frac{2 \pi}{3}\left( \frac{3 r^{6}_{H}-8 r^{2}_{H}q^{4}}{8q^{4}+ r^{4}_{H}}   \right)+\left\{-\frac{2^{5/2}q}{\pi q^{3}}\left[\frac{ 2r_{H}+2^{5/2}q}{r^{2}_{H}+2^{5/2}qr_{H}+2\sqrt{2}}-\frac{2r_{H}-2^{5/2}q}{r^{2}_{H}-2^{5/2}qr_{H}+2\sqrt{2}}    \right] \right.\\
&\left.-4\left[\frac{1}{1+\left( \frac{r_{H}}{2^{1/4}q}+1 \right)^{2}}  +\frac{1}{1+\left( \frac{r_{H}}{2^{1/4}q}-1 \right)^{2}} \right]-\frac{16q^{4}}{ r^{4}_{H}}-\frac{16 \pi q^{4}}{ r^{2}_{H}} \right\}\left(\frac{r^{4}_{H}}{3(8q^{4}+r^{4}_{H})}   \right).
\end{aligned}
\end{equation}

\begin{figure}[h]
  \centering
  \includegraphics[scale=0.6]{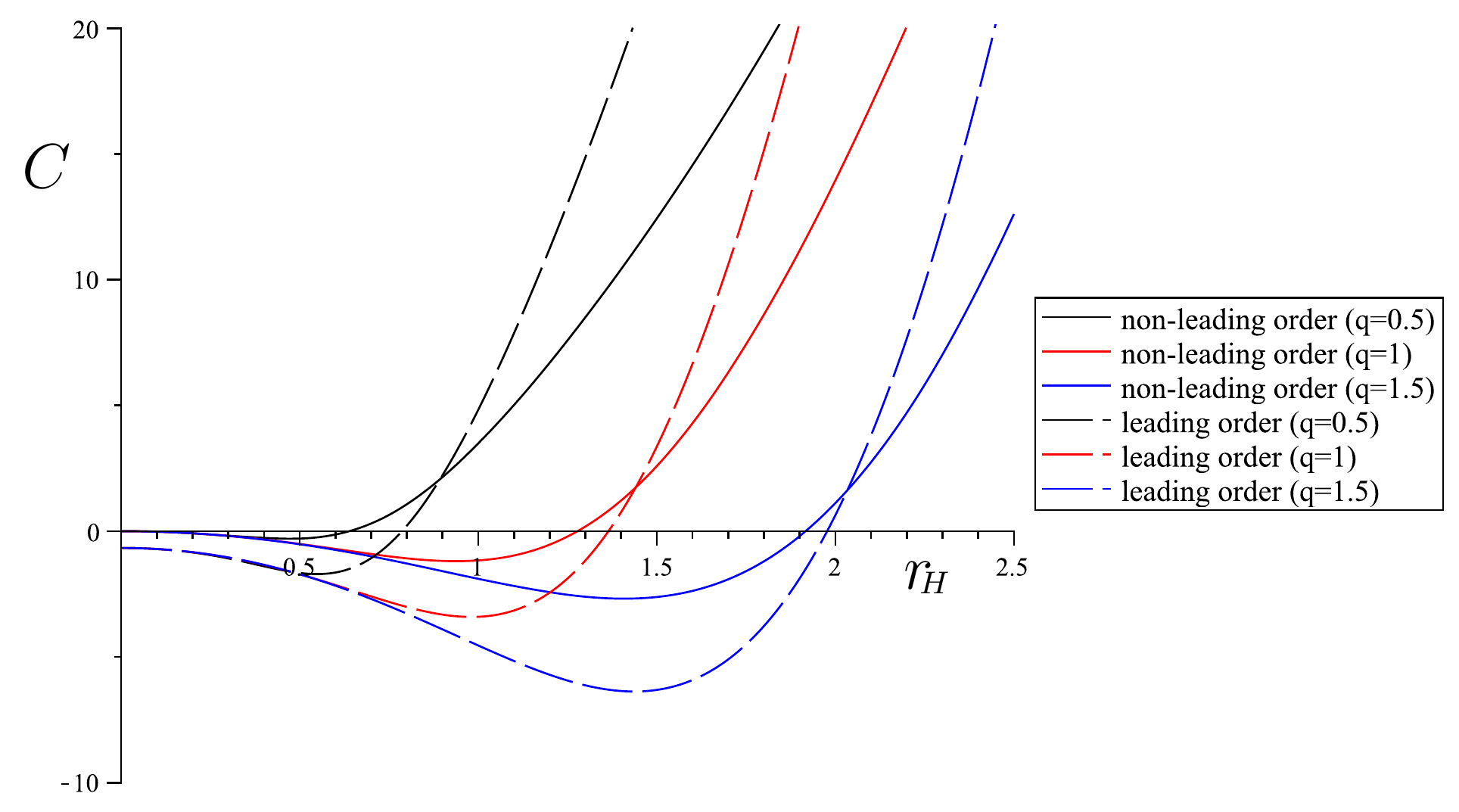}
  \caption{Corrected specific heat versus the black hole horizon for various $q$ values.} \label{fig8}
\end{figure}
\newpage

{\color{black}To analyze the behavior of the specific heat and the effects of thermal fluctuations, we plot the graph shown in Fig. \eqref{fig8}. From the figure, we can see that the specific heat is almost zero for a NLED black hole in thermal equilibrium. However, when the small statistical fluctuations around equilibrium are considered, we obtain a non-zero specific heat. Moreover, the graph shows that the specific heat generally increases as the radius of the horizon increases. For small black holes, the specific heat has negative values, but it can get more negative with thermal fluctuations. This shows that thermal fluctuations cause the black holes to enter an unstable phase earlier. On the other hand, large (having bigger $r_{H}$ ones) NLED black holes are thermally stable.}

\section{Quantum Wave Probe in Static Spacetimes} \label{sec4}
The method proposed in \citep{8,9}, translates the dynamics of the scalar wave evolution into the problem of finding a unique self-adjoint extension to the spatial part of the Hamiltonian wave operator ( Reader is refereed to \citep{16,17,18,19}) for a detailed mathematical background of the subject). This method was originally developed for scalar waves obeying the massive Klein - Gordon equation that can be written by splitting temporal and spatial parts:
\begin{equation}
\frac{\partial ^{2}\Psi }{\partial t^{2}}=-A\Psi, \label{s51}
\end{equation}%
in which $A=-VD^{i}\left( VD_{i}\right)+V^{2}m^{2} $ is a spatial operator containing spatial coordinates with $
V^{2}=-\xi _{\mu }\xi ^{\mu }$ and $m$ denotes the mass. In this formalism, $\xi ^{\mu }$ stands for a
time-like Killing field and $D_{i}$ is the spatial covariant derivative on a
static slice $\Sigma .$ The suitable function space is the usual Hilbert space $\mathcal{H}$, characterized by $L^{2}\left( \Sigma  \right) .$  It is worth to emphasize that the scalar field dynamics can be generalized to other fields such as Dirac  (spin-$1/2$) and Maxwell fields (spin-$1$) \citep{10,11,16,18}. \\
The crucial point in this method is to show that the self-adjoint extension of the spatial operator $ A_{E}$
is unique, which is said to be essentially self-adjoint. When this is the case,
the Klein-Gordon equation for a free relativistic scalar particle satisfies%
\begin{equation}
i\frac{d\Psi }{dt}=\sqrt{A_{E}}\Psi, \label{s52}
\end{equation}%
whose solution reads
\begin{equation}
\Psi \left( t\right) =\exp (-it\sqrt{A_{E}})\Psi \left( 0\right). \label{s53}
\end{equation}%
If the spatial operator $A_{E}$ has more than one self-adjoint extension, there will be more than one future time evolution of the wave function as defined in Eq. \eqref{s53}. This result is ambiguous and the spacetime is said to be quantum singular. However, if the operator $A_{E}$ has a unique self-adjoint extension, then Eq.\eqref{s53}  implies unique evolution and in this case, the spacetime is said to be quantum regular. \\
Studying quantum singularities in static spacetimes using the method in  \citep{8,9}, requires counting the number of self-adjoint extensions of the spatial wave operator $A$. The well-known theory in this direction is the concept of deficiency indices discovered by Weyl \citep{20} and generalized by von Neumann \citep{21}. According to the von Newmann theorem, the deficiency indices are determined by counting the number of solutions of $(A\pm i)\psi=0$ (for each sign of $i$) that belongs to the Hilbert space $\mathcal{H}$. Solutions that are not square integrable do not belong to the Hilbert space, and hence, the deficiency indices are $n_{+}=n_{-}=0$. In this case, it is said that the spatial operator $A$ is essentially self-adjoint. Otherwise, if $n_{+}=n_{-}=n\geq 1$, in other words, if the solutions are square integrable, this indicates that the spatial operator $A$ has many self-adjoint extensions. \\
\subsection{Scalar Field Probe}
The massive Klein-Gordon wave equation $(\nabla^{\mu} \nabla_{\mu}-m^{2})\psi=0 $ can be written by splitting time and spatial part for the metric \eqref{s1} as

\begin{equation}
\begin{aligned}
\frac{\partial ^{2} \psi}{\partial t^{2}}=&f^{2}(r)\frac{\partial ^{2} \psi}{\partial r^{2}}+\frac{f(r)}{r^{2}}\frac{\partial ^{2} \psi}{\partial \theta^{2}}+\frac{f(r)}{r^{2}sin^{2}\theta}\frac{\partial ^{2} \psi}{\partial \varphi^{2}}+\frac{f(r)cot \theta}{r^{2}}\frac{\partial  \psi}{\partial \theta}\\
 &+f(r)\left ( \frac{2f(r)}{r}+f^{'}(r)\right)\frac{\partial  \psi}{\partial r}-f(r)m^{2}\psi.
\end{aligned} \label{s55}
\end{equation}

The spatial Hamiltonian wave operator can be defined by comparing the above equation with Eq.\eqref{s51}, which reads

\begin{equation}
\begin{aligned}
A=&-f^{2}(r)\frac{\partial ^{2} }{\partial r^{2}}-\frac{f(r)}{r^{2}}\frac{\partial ^{2} }{\partial \theta^{2}}-\frac{f(r)}{r^{2}sin^{2}\theta}\frac{\partial ^{2} }{\partial \varphi^{2}}-\frac{f(r)cot \theta}{r^{2}}\frac{\partial }{\partial \theta} \\
&-f(r)\left ( \frac{2f(r)}{r}-f^{'}(r)\right)\frac{\partial}{\partial r}+f(r)m^{2}.
\end{aligned} \label{s56}
\end{equation}

By using the separation of the variables $ \psi =R(r)Y_{l}^{m}( \theta , \varphi )$, the radial part of the equation $(A\pm i)\psi=0$ becomes

\begin{equation}
R^{''}+\frac{[r^{2}f(r)]^{'}}{r^{2}f(r)}  R^{'}+\left [ \frac{-l(l+1)}{r^{2}f(r)}-\frac{  m^{2}}{f(r)} \mp \frac{i}{f^{2}(r)}\right]R=0. \label{s57}
\end{equation}

The square integrability condition of solutions \eqref{s57} for each sign $ \pm$ is checked by calculating the squared norm, given by \citep{9}

\begin{equation}
||R||^{2}=\int_{\Sigma _{t}}\sqrt{-g}g^{tt}RR^{*}d^{3}\Sigma _{t}. \label{s58}
\end{equation}

The square integrability of the solutions provided by Eq.\eqref{s57} will be analyzed for the whole space. By doing so, the behavior of Eq.\eqref{s57}  near $r\rightarrow 0$ and $r \rightarrow \infty$ will be studied separately. \\
In the case when $r\rightarrow 0$, the metric function \eqref{m3} behaves as
\begin{equation}
f(r)\approx 1+\sqrt{\eta-\beta r} \label{s59}
\end{equation}%

in which $\beta=\frac{4M \tilde{\Lambda} }{3}$, $\eta=\frac{2q^{2} \tilde{\Lambda}}{3}$, and $ \tilde{\Lambda}>0$. When $ r \rightarrow 0 $, Eq.\eqref{s57} reduces to

\begin{equation}
R^{''}+ \frac{2}{r}R^{'}-\frac{l(l+1)}{(1+\sqrt{\eta}) r^{2}} R =0, \label{s60}
\end{equation}

The solution of this equation is given by

\begin{equation}
R(r)=c_{1}r^{\frac{(-1+\sqrt{1+4a})}{2}}+c_{2}r^{\frac{(-1-\sqrt{1+4a})}{2}}, \label{s61}
\end{equation}

with $a=\frac{l(l+1)}{1+\sqrt{\eta}}$ and $c_{1},c_{2}$ are integration  constants. Substituting Eq.\eqref{s61} into Eq.\eqref{s58}, one finds

\begin{equation}
\begin{aligned}
\| R \|^{2} \sim   \int_{const.}^{0}\frac{r^{2}|R|^{2}}{1+\sqrt{\eta -\beta r }}dr&=c_{1}^{2}\int_{const.}^{0}\frac{r^{1+p}}{1+\sqrt{\eta -\beta r }}dr+2c_{1}c_{2}\int_{const.}^{0}\frac{r}{1+\sqrt{\eta -\beta r }}dr \\
&
+c_{2}^{2}\int_{const.}^{0}\frac{r^{1-p}}{1+\sqrt{\eta -\beta r }}dr,
\end{aligned} \label{s62}
\end{equation}

where $p=\sqrt{1+4a}$. The first and last terms of Eq.\eqref{s62}  can be investigated by using the comparison test. As a result, one can comment on whether the integral diverges or converges. Applying the comparison test to the first integral leads to the inequality

\begin{equation}
0 \leq \frac{r^{1+p}}{1+\sqrt{\eta -\beta r }} \leq     \frac{r^{1+p}}{\sqrt{\eta -\beta r  }}.
\end{equation}
Once the integral of $ r^{1+p}/\sqrt{\eta -\beta r  }$ is calculated, it gives

\begin{equation}
\int_{const.}^{0}\frac{r^{1+p}}{\sqrt{\eta -\beta r  }}dr=\frac{1}{\eta}\left( 2r^{2+p} \sqrt{\eta-\beta r}\right) {}_{2}F_{1}\left( 1,p+\frac{5}{2};\frac{3}{2};1-\frac{\beta r}{\eta} \right)  |_{const}^{0}<\infty,
\end{equation}
where ${}_{2}F_{1}\left( 1,p+\frac{5}{2};\frac{3}{2};1-\frac{\beta r}{\eta} \right)$ is the hypergeometric function. As the upper limit is found to be convergent as a consequence of the comparison test, one can conclude that the first integral is also convergent. In other words, the solution to the first integral is finite, and in turn, belongs to the Hilbert space. Following similar steps, the second integral can also be checked for the space that it belongs to. The evaluations allow one to express the second integral in the form

\begin{equation}
\begin{aligned}
2c_{1}c_{2}\int_{const.}^{0}\frac{r}{1+\sqrt{\eta -\beta r }}dr =\frac{2c_{1}c_{2}}{3\beta^{2}}&\left \{ 2(3-\eta)\sqrt{\eta -\beta r }-6(\eta-1)ln\left(  \sqrt{\eta -\beta r }+1  \right)   \right.\\
 &\left.+ \beta r \left( 3-\sqrt{\eta - \beta r} \right) \right \} |_{const.}^{0} <\infty,
 \end{aligned}
\end{equation}
which implies square integrability. The convergence character of the last integral can be analyzed by using the comparison test. With this purpose, the following inequality is studied

\begin{equation}
0 \leq \frac{r^{1-p}}{1+\eta -\beta r } \leq     \frac{r^{1-p}}{1+\sqrt{\eta -\beta r }},
\end{equation}

The integral of $\frac{r^{1-p}}{(1+\eta) -\beta r}$ is calculated as

\begin{equation}
\int_{const.}^{0}\frac{r^{1-p}}{1+\eta-\beta r}dr=\frac{r^{2-p}{}_{2}F_{1}\left( 1,2-p;3-p;\frac{\beta r}{1+\eta}    \right)}{(1+\eta)(2-p)} |_{const}^{0},\label{m11}
\end{equation}

where ${}_{2}F_{1}\left( 1,2-p;3-p;\frac{\beta r}{1+\eta}\right)$ is the hypergeometric function.\\
Since ${}_{2}F_{1}\left( 1,2-p;3-p;0\right)=1$  and $3 \leq p$, integral \eqref{m11}  diverges. According to the comparison test, the divergence of integral $\int_{const.}^{0}\frac{ r^{1-p}}{1+\eta -\beta r} dr$,  implies the divergence of $\int_{const.}^{0} \frac{r^{1-p}}{1+\sqrt{\eta -\beta r }}dr$.

The square integrability analysis has revealed that whenever $2 \leq\frac{ l(l+1)}{1+\sqrt{\eta}}$, the squared norm diverges. This implies that the solution for this particular choice fails to be square integrable. However, if $2>\frac{l(l+1)}{1+\sqrt{\eta}}$, then the solution is square integrable.

In the case when $ r \rightarrow \infty $, the metric function can be approximated as

\begin{equation}
f(r)\approx 1+\delta r^{2},\label{m20}
\end{equation}%

in which $\delta=\tilde{\Lambda}/3$. Eq.\eqref{s57}  reduces to

\begin{equation}
R^{''}+ \frac{4}{r}R^{'} =0,
\end{equation}

whose solution is given by

\begin{equation}
R(r)=c_{3}+\frac{c_{4}}{r^{3}},\label{m12}
\end{equation}
where $c_{3},c_{4}$  are  the  integration  constants. Substituting Eq.\eqref{m12}  into Eq.\eqref{s58}  leads to

\begin{equation}
\begin{aligned}
\| R \|^{2} \sim  \int_{const.}^{\infty}\frac{r^{2}|R|^{2}}{1+\delta r^{2}}dr=&c_{3}^{2}\int_{const.}^{\infty}\frac{r^{2}}{1+\delta r^{2}}dr+2c_{3}c_{4}\int_{const.}^{\infty}\frac{1}{r(1+\delta r^{2})}dr\\
&+c_{4}^{2}\int_{const.}^{\infty}\frac{1}{r^{4}(1+\delta r^{2})}dr \label{m13}
\end{aligned}
\end{equation}
The integration of the first integral reveals

\begin{equation}
c_{3}^{2}\int_{const.}^{\infty}\frac{r^{2}}{1+\delta r^{2}}dr=c_{3}^{2}\left [\frac{r}{\delta}-\frac{tan^{-1}(\sqrt{\delta}r)}{\delta^{3/2}} \right ]  |_{const}^{\infty} \rightarrow \infty,\label{m14}
\end{equation}
whereas the second integral is found to be
\begin{equation}
2c_{3}c_{4}\int_{const.}^{\infty}\frac{1}{r(1+\delta r^{2})}dr=2c_{3}c_{4}\left [ ln(r)-\frac{1}{2}ln(\delta r^{3} +1) \right]    |_{const}^{\infty} \rightarrow \infty. \label{m15}
\end{equation}
Finally, the last integral results in

\begin{equation}
c_{4}^{2}\int_{const.}^{\infty}\frac{1}{r^{4}(1+\delta r^{2})}dr=c_{4}^{2}\left [ \frac{\delta}{r}-\frac{1}{3r^{3}}+\delta^{3/2}tan^{-1}(\sqrt{\delta}r) \right]    |_{const}^{\infty} < \infty.\label{m16}
\end{equation}

Having evaluated all the integrands in Eq.\eqref{m13}  analytically, we can now comment on the final form of the square norm. Gathering Eqs.\eqref{m14}-\eqref{m16} together, it is found that the integrability condition results in $||R||^{2} \rightarrow \infty$ . All these calculations indicate that there exist specific solutions of Eq.\eqref{s57}, near $ r \rightarrow 0 $ and $ r \rightarrow \infty $, that are not square integrable. In other words, $n_{+}=n_{-}=0$ . This shows that the spatial operator $A$ is uniquely defined (essentially self-adjoint) for the entire space. Physically, the future time evolution of the quantum wave packets obeying the Klein-Gordon equation is uniquely determined. Consequently, the classical time-like naked singularity in the space of the considered spacetime of $\mathcal{F}(R,\mathcal{G})$ gravity becomes quantum mechanically regular.

\subsection{ Probe with Dirac fields }
Chandrasekhar - Dirac (CD) equations in the Newman - Penrose formalism are given by  \citep{22}, 

\begin{eqnarray}
\left( D+\epsilon -\rho \right) F_{1}+\left( \bar{\delta}+\pi -\alpha
\right) F_{2} &=&0, \label{is75}\\
\left( \Delta +\mu -\gamma \right) F_{2}+\left( \delta +\beta -\tau \right)
F_{1} &=&0,  \notag \\
\left( D+\bar{\epsilon}-\bar{\rho}\right) G_{2}-\left( \delta +\bar{\pi}-%
\bar{\alpha}\right) G_{1} &=&0,  \notag \\
\left( \Delta +\bar{\mu}-\bar{\gamma}\right) G_{1}-\left( \bar{\delta}+\bar{%
\beta}-\bar{\tau}\right) G_{2} &=&0,  \notag 
\end{eqnarray} 
where $F_{1},F_{2},G_{1}$, and $G_{2}$ are the components of the wave
function, $\epsilon ,\rho ,\pi ,\alpha ,\mu ,\gamma ,\beta $ and $\tau $ are
the spin coefficients. The solution procedure of the set of CD equations \eqref{is75} will be ignored in this study since it will be a repetition of references \citep{10,11}. We prefer to give only the final answer in the form of  one-dimensional Schr\"{o}dinger-like wave equation
with an effective potential that governs the Dirac field,

\begin{gather}
\left( \frac{d^{2}}{dr_{\ast }^{2}}+k^{2}\right) Z_{\pm }=V_{\pm }Z_{\pm },
\\
V_{\pm }=\left[ \frac{f\lambda ^{2}}{r^{2}}\pm \lambda \frac{d}{dr_{\ast }}%
\left( \frac{\sqrt{f}}{r}\right) \right] .
\end{gather}%
In these equations, $Z_{\pm }=R_{1}\pm R_{2}$ are the two solutions of the CD equations and $\lambda $ denotes the
separability constant. With the help of Eq.\eqref{s51}, the radial operator $A$ for the Dirac equations can be rewritten as

\begin{equation*}
A=-\frac{d^{2}}{dr_{\ast }^{2}}+V_{\pm }.
\end{equation*}%
The radial operator is written in terms of the usual radial coordinate $r$ by
using $\frac{d}{dr_{\ast }}=f\frac{d}{dr}$, we then obtain

\begin{equation}
A=-\frac{d^{2}}{dr^{2}}-\frac{f^{^{\prime }}}{f}\frac{d}{dr}+\frac{1}{f^{2}}%
\left[ \frac{f\lambda ^{2}}{r^{2}}\pm \lambda f\frac{d}{dr}\left( \frac{%
\sqrt{f}}{r}\right) \right] .
\end{equation}

From this point onward, one needs to check whether this radial part of the Dirac operator
is essentially self-adjoint. This can be done by considering equation $(A\pm i)\psi=0$ and
counting the number of solutions that do not belong to Hilbert space. Thus,
we have,

\begin{equation}
\left( \frac{d^{2}}{dr^{2}}+\frac{f^{^{\prime }}}{f}\frac{d}{dr}-\frac{1}{%
f^{2}}\left[ \frac{f\lambda ^{2}}{r^{2}}\pm \lambda f\frac{d}{dr}\left(
\frac{\sqrt{f}}{r}\right) \right] \mp i\right) \psi (r)=0.\label{m17}
\end{equation}%
The solutions of Eq.\eqref{m17}  should be checked for square integrability over the entire
space $L^{2}\left( 0,\infty \right)$. In other words, the solutions of Eq.\eqref{m17}  will be analyzed near $ r \rightarrow 0 $ and $ r\rightarrow \infty,$  for the essential self-adjointness of the spatial operator. \\
In the case when $ r \rightarrow 0,$ the approximate metric function is obtained in Eq.\eqref{s59} and
using this equation in Eq.\eqref{m17} yields
\begin{equation}
\psi ^{\prime \prime }-\frac{\beta}{2 \bar{\eta}}\psi ^{\prime }-\frac{\bar{\lambda}^{2} }{r^{2}}%
\psi =0,
\end{equation}%
where $\bar{\eta}=\eta+\sqrt{\eta}$ and $\bar{\lambda}^{2}=\frac{\lambda^{2}}{1+\sqrt{\eta}},$
and the solution is given by%
\begin{equation}
\psi (r)=c_{5}e^{\alpha r}\sqrt{r}J_{\sigma}\left( \frac{\alpha i}{4}r\right) +c_{6}e^{\alpha r}\sqrt{r}Y_{\sigma}\left(
\frac{\alpha i}{4}r\right).
\end{equation}%
Note that $c_{5}$, $c_{6}$\ are integration constants and $J$ and $Y$ are the Bessel function of the first kind and the Bessel function of the second kind, respectively with $\alpha=\frac{\beta}{4\bar{\eta}}$ and $\sigma=\sqrt{1+4\bar{\lambda}}/2$. The behaviour of the Bessel functions for real $\nu \geq 0$ as $ r \rightarrow 0 $ are defined by \citep{23}

\begin{equation}
\begin{aligned}
J_{\nu}(r) &\sim \frac{1}{\Gamma(\nu +1)}\left ( \frac{r}{2}\right )^{\nu}\\
Y_{\nu}(r) &\sim \begin{cases}
\frac{2}{\pi}\left[ ln \left ( \frac{r}{2} \right ) +\gamma \right ]&,\nu =0 \;and \; \gamma \cong 0.5772\\
-\frac{\Gamma(\nu)}{\pi}\left(\frac{2}{r}\right)^{\nu}&,\nu\neq 0 \: \: \: \: \: \: \: \: \:\: \: \: \: \: \: \: \: \: \: \: \: \: \: \: \: \: \: \: \: \: \: \:
\end{cases}
\end{aligned}
\end{equation}

The solution then can be written as

\begin{equation}
\psi (r) \cong -  \bar{c_{6}} r^{\frac{1-2\sigma}{2}}, \label{m18}
\end{equation}

with $\bar{c_{6}}=\frac{C_{6}\Gamma(\sigma)8^{\sigma}}{\pi (i\alpha)^{\sigma }}$. Once Eq.\eqref{m18} is substituted into the squared norm \eqref{s58},  it becomes

\begin{equation}
\| R \|^{2} \sim  \int_{const.}^{0}\frac{r^{2}|R|^{2}}{1+\sqrt{\eta-\beta r}}dr=\bar{c_{6}}^{2}  \int_{const.}^{0}\frac{r^{3-2\sigma}}{1+\sqrt{\eta-\beta r}}dr.
\end{equation}

 The convergence character of the integral is analyzed by using the comparison test as before. We define the required inequality for the comparison test as:

\begin{equation}
0 \leq \frac{r^{3-2\sigma}}{1+\eta -\beta r } \leq     \frac{r^{3-2\sigma}}{1+\sqrt{\eta -\beta r }}.
\end{equation}
The integral of $\frac{r^{3-2\sigma}}{(1+\eta) -\beta r}$ can be calculated as

\begin{equation}
\int_{const.}^{0}\frac{r^{3-2\sigma}}{1+\eta -\beta r}dr=\frac{r^{4-2\sigma}{}_{2}F_{1}\left( 1,4-2\sigma;5-2\sigma;\frac{\beta r}{1+\eta}    \right)}{(1+\eta)(4-2\sigma)} |_{const}^{0}, \label{m19}
\end{equation}

where ${}_{2}F_{1}\left( 1,4-2\sigma;5-2\sigma;\frac{\beta r}{1+\eta}    \right)$ is the hypergeometric function.\\
Since ${}_{2}F_{1}\left( 1,4-2\sigma;5-2\sigma;\frac{\beta r}{1+\eta}    \right)=1$  and $5 \leq 2\sigma$, integral \eqref{m19} diverges. According to the comparison test, the divergence of integral $\int_{const.}^{0}\frac{ r^{1-p}}{1+\eta -\beta r} dr$,  imposes the divergence of $\int_{const.}^{0} \frac{r^{1-p}}{1+\sqrt{\eta -\beta r }}dr$. The square integrability analysis has revealed that whenever $6 \leq \bar{\lambda}^{2}$, the squared norm diverges. This implies that the solution for this particular choice fails to be square integrable. However, if $6>\bar{\lambda}^{2}$, the solution is square integrable.

In the case when $r\rightarrow \infty $, the metric function \eqref{m20} is used in \eqref{m17}, which becomes

\begin{equation}
\psi^{''}+ \frac{2}{r}\psi^{'} =0,
\end{equation}

and the solution is given by
\begin{equation}
R(r)=c_{7}+\frac{c_{8}}{r},\label{m21}
\end{equation}

in which $c_{7},c_{8}$  are  the  integration  constants. Substituting Eq.\eqref{m21}  into Eq.\eqref{s58}  leads to

\begin{equation}
\begin{aligned}
\| R \|^{2} \sim  \int_{const.}^{\infty}\frac{r^{2}|R|^{2}}{1+\delta r^{2}}dr=&c_{7}^{2}\int_{const.}^{\infty}\frac{r^{2}}{1+\delta r^{2}}dr+2c_{7}c_{8}\int_{const.}^{\infty}\frac{r}{1+\delta r^{2}}dr\\
&+c_{8}^{2}\int_{const.}^{\infty}\frac{1}{1+\delta r^{2}}dr.
\end{aligned}
\end{equation}

The first integral gives

\begin{equation}
c_{7}^{2}\int_{const.}^{\infty}\frac{r^{2}}{1+\delta r^{2}}dr=c_{7}^{2}\left [\frac{r}{\delta}-\frac{tan^{-1}(\sqrt{\delta}r)}{\delta^{3/2}} \right ]  |_{const}^{\infty} \rightarrow \infty.
\end{equation}

The result of second integral is found as
\begin{equation}
2c_{7}c_{8}\int_{const.}^{\infty}\frac{r}{1+\delta r^{2}}dr=c_{7}c_{8}\left [ \frac{1}{\delta}ln(\delta r^{2} +1) \right]    |_{const}^{\infty} \rightarrow \infty,
\end{equation}
and the last integral is solved as

\begin{equation}
c_{4}^{2}\int_{const.}^{\infty}\frac{1}{1+\delta r^{2}}dr=c_{8}^{2}\left [ \sqrt{\delta}tan^{-1}(\sqrt{\delta}r) \right]    |_{const}^{\infty} < \infty.
\end{equation}

From these calculations, it is clear that the squared norm diverges,  $||R||^{2} \rightarrow \infty$. Since the specific solutions are not square integrable, the corresponding spatial Dirac operator becomes essentially self-adjoint. As a result, the timelike naked singularity in the considered spacetime remains quantum mechanically regular with respect to specific modes of quantum wave packets containing fermionic fields.

\section{Conclusions} \label{sec5}

Understanding the underlying physics of the extended theories of Einstein's classical theory of general relativity is extremely important. In this study, topological static spherically symmetric solutions in $\mathcal{F}(R,\mathcal{G})$-gravity coupled with Born- Infeld like NLED is investigated in terms of the tools of quantum mechanics. Thermal properties are investigated by considering corrected thermodynamic variables. On the other hand, solutions admitting timelike naked singularities are investigated by probing the singularity with quantum wave packets. It is worth emphasizing here that there are c  number of interesting solutions in the literature that involves nonlinear electrodynamics fields in different form of Lagrangian \cite{24i,25i,26i,27i,28i,29i,31i,32i,33i}. Analysis of those solutions would be an interesting task an our future studies along this direction. 

{\color{black} We have looked into how statistical fluctuations affect the NLED black holes. We have discovered that the entropy of charged static NLED black holes is (logarithmically) corrected as a result of thermal fluctuations near the equilibrium. To discuss the impact of thermal fluctuations on the entropy of a non-rotating NLED black hole, we have generated a graph of entropy versus horizon radius (see Fig. \eqref{fig2}). We have found that there is a critical value below which the corrected entropy becomes a decreasing function and above which the inverse is true. The corrected entropy, however, never experiences negative values, as it should. With the aid of the first law of thermodynamics, we have also evaluated the corrected total energy (mass) values for a NLED black hole and discovered that the total mass of the system exhibits an exponential type function (see Fig. \eqref{fig3}) with respect to the event horizon radius. The system's overall mass is reduced by the thermal fluctuation-related correction term. Further, we have determined the first-order corrected Helmholtz free energy for a NLED black hole where the corrections appear due to thermal fluctuations. By analytically graphing the uncorrected and corrected free energies, we have examined the Helmholtz free energy's behavior and conducted a comparison analysis (see Fig. \eqref{fig4}). We have discovered that the Helmholtz free energy decreases with horizon radius. The free energy becomes more positive with the inclusion of correction terms. The less charge $q$ parameters, however, make the free energy more negative for the smaller NLED black holes.

Moreover, we have determined the system's corrected pressure. According to Fig. \eqref{fig5}, the corrected pressure for massive black holes aligns with the equilibrium pressure and reaches saturation. The pressure takes asymptotically large values for very small black holes because of the thermal fluctuations. The minor thermal changes do not, in fact, have the expected noticeable effects on the pressure for the big NLED black holes. 

We have also determined the enthalpy's corrected expression. We have deduced from the associated  plots (Fig. \eqref{fig6n}) that the enthalpy behaves as if the total energy. The enthalpy value is raised by thermal variations. In other words, the enthalpy grows as the correction parameter's value increases. For the NLED black holes with lower charges, the Gibbs free energy might be negative, while for those with higher charges, it is usually positive (see Fig. \eqref{fig7}). For smaller NLED black holes, the corrected Gibbs free energy with a bigger charge parameter gets asymptotic positive values.

The specific heat becomes an increasing function with respect to the event horizon radius. As can be seen from Fig. \eqref{fig8}, the specific heat is negative for low horizon radii of the NLED black holes. However, it quickly becomes positive with increasing radius, which suggests that NLED black holes are in a stable phase devoid of thermal fluctuations. For small black holes, the specific heat becomes negative because of the thermal oscillations. This shows that small black holes experience temperature fluctuations and are in an unstable phase. Besides, if the charge parameter is increased, bigger NLED black holes become more stable.}
 \\
When the classical singularity is probed with particular modes of quantum wave packets obeying Klein-Gordon and Dirac equations, it is demonstrated that the formation of classical timelike naked singularity becomes quantum mechanically regular. Extensive calculations have shown that whenever the mode of the quantum wave packet satisfies the condition $2 \leq \frac{l(l+1)}{1+\sqrt{\eta}}$, the spatial wave operator for the Klein-Gordon fields becomes basically self-adjoint. The spatial wave operator's self-adjointness of the Dirac fields is similarly subject to similar restrictions. When a fermionic quantum wave packet's mode satisfies the condition $\overline{\lambda}\geq 6 $, the corresponding wave operator  becomes essentially self-adjoint. These results indicates that there some modes of solutions such that the corresponding Hamiltonian wave operator turns out to be essentially self-adjoint, which in turn implies unique well-defined time evolution and thus it can be said that the classically singular spacetime becomes quantum mechanically regular. \\
Due to the deterministic character of the theory of general relativity, it is crucial to comprehend and eliminate the curvature singularities in classical general relativity. If a reliable theory of quantum gravity is developed, we think this issue is going to be solved. Future studies can be made to focus on the appearance of timelike naked singularities in modified theories in light of quantum mechanics. Thus, one can examine whether these singularities may be cured by quantum effects or not. These topics will be on our upcoming job list.

\section*{Acknowledgement}
{\color{black}We would like to express our gratitude to the Editor and Reviewers, for overseeing the review process and providing valuable guidance. Your efforts in coordinating the reviews and ensuring the integrity of the publication process are truly commendable. We extend our special thanks to Professor Elias C. Vagenas, who identified and shared the typos in his work \cite{6}, and facilitated highly productive scientific discussions.} We also acknowledge the contributions of T\"{U}B\.{I}TAK, ANKOS, and SCOAP3. Furthermore, \.{I}.S. would like to acknowledge networking support of COST Actions CA22113, CA18108, and CA21106.

\end{document}